\documentclass[11pt]{article}
\usepackage{moriond}

\usepackage{graphicx,amssymb,bbm,mathrsfs,cite,epstopdf}
\pdfoutput=1

\def\met{\mbox{${\hbox{$E$\kern-0.6em\lower-.1ex\hbox{/}}}_T$}}

\newcommand {\unit} [1] {\; \mathrm {#1}}

\def\ti              {\tilde}
\def\x               {\chi}

\def\ch              {\ti \x^\pm}

\def\nt              {\ti \x^0}

\def\be{\begin{equation}}
\def\ee{\end{equation}}
\def\bea{\begin{eqnarray}}
\def\eea{\end{eqnarray}}

\def\fbi {~fb$^{-1}$}
\begin{document}
\vspace*{4cm}
\title{SUSY STATUS AFTER ONE YEAR OF LHC}

\author{ S. KRAML }

\address{Laboratoire de Physique Subatomique et de Cosmologie, UJF Grenoble 1, CNRS/IN2P3, INPG, \\
53 Avenue des Martyrs, F-38026 Grenoble, France}

\maketitle\abstracts{
I review the status of supersymmetry after the 2011 LHC search results. I concentrate in particular on interpretations beyond the conventional CMSSM, including {\it i)} natural SUSY with light stops, and {\it ii)}  the so-called phenomenological MSSM, which is a general parametrization of the MSSM at the weak scale, without boundary conditions imposed by specific SUSY breaking schemes. We will see that the current searches are not yet sensitive to some of the theoretically most interesting scenarios.}

\section{Introduction}

\begin{figure}[t]\centering
\includegraphics[width=10cm]{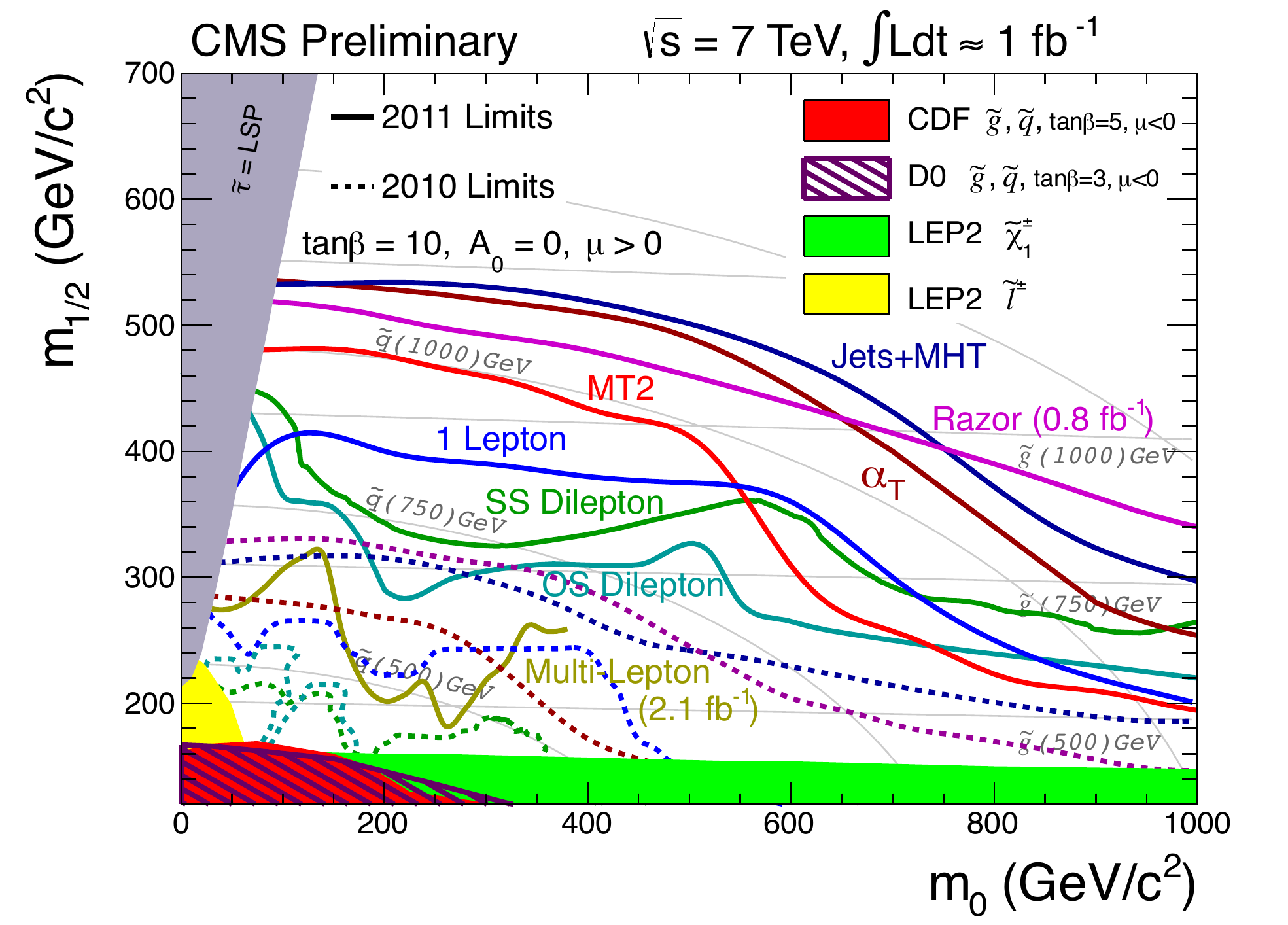}  
\caption{ \label{fig:CMSSMLimit}
Observed limits from several 2011 CMS SUSY searches plotted in the CMSSM, from \cite{cms:twiki}. }
\end{figure}

Before the LHC turn-on, we had a very optimistic view:  if SUSY is light (as we of course all expected) it will be discovered early on, in particular much earlier than the Higgs, which is much harder to find. Today, with about 5\fbi of data per experiment collected at $\sqrt{s}=7$~TeV,\footnote{And, at the time of writing, more data coming and being analyzed at $\sqrt{s}=8$~TeV.} the situation is quite different. Tantalizing hints of a Higgs signal are emerging \cite{higgshints}, but there is no signal of new physics whatsoever. Indeed, the direct search limits from ATLAS and CMS are pushing the SUSY mass scale to $M_{\rm SUSY} > 1$~TeV~\cite{lowette,atlas:twiki,cms:twiki}, at least in simplistic models like the constrained MSSM (CMSSM), see Fig.~\ref{fig:CMSSMLimit}. 
Even worse, precision measurements in the flavor sector are pushing the (flavor) scale of new physics out into the multi-TeV range~\cite{flavor}. 

So is SUSY in trouble? To assess this question, let us first (re)consider the main arguments why we generally like to expect $M_{\rm SUSY} < 1$~TeV. 

\begin{itemize}

\item A solution to the naturalness and gauge hierarchy problems: this relies on the cancellation of quadratic divergencies of the Higgs mass, when embedded in e.g.\ a GUT theory, see e.g.~\cite{Martin:1997ns}. For this the superpartners of the particles which have a large coupling to the Higgs should be not too heavy. Therefore we expect in particular light stops, but also light higgsinos, and a somewhat light gluino (to keep the stop light). The rest of the spectrum may however be heavier, above a TeV.

\item Gauge coupling unification: this needs first of all new TeV-scale fermionic states (in addition to a light Higgs doublet). In the MSSM  this means light gauginos and higgsinos. The scalars could be heavy---even superheavy like in split SUSY~\cite{ArkaniHamed:2004fb,Giudice:2004tc}

\item Radiative electroweak symmetry breaking (REWSB) and a light Higgs~\cite{Martin:1997ns}: to my mind an extremely attractive feature of the MSSM. However, once the particle content of the model is supersymmetric, REWSB is essentially is a heavy top effect, independent of the exact values of the soft-breaking terms. Besides, $m_h>115$ GeV prefers somewhat heavy stops (the so-called finetuning prize of LEP~\cite{Chankowski:1997zh,Barbieri:1998uv}), which will get only more severe if $m_h\simeq 125$~GeV is confirmed, see e.g.~\cite{Brummer:2012ns}. 
Furthermore, electroweak (EW) precision measurements prefer heavy SUSY~\cite{Heinemeyer:2006px}.

\item Dark matter candidate: if R-parity is conserved, the lightest SUSY particle (LSP) is an excellent dark matter candidate. 
However, a TeV-scale LSP could very well do the job, it just needs some efficient annihilation mechanism. Indeed, a pure higgsino LSP, for instance, should weight about 1 TeV to have $\Omega h^2\simeq 0.1$.


\end{itemize}

Another very important point to keep in mind is that most SUSY mass limits have been obtained within the CMSSM, 
which is characterized by just four-and-a-half parameters: 
a universal scalar mass $m_0$, 
gaugino mass $m_{1/2}$ and trilinear coupling $A_0$  
defined at the GUT scale $M_{\rm GUT}\sim 10^{16}$~GeV, plus 
$\tan\beta$ and $\rm{sign(\mu)}$. 
The complete spectrum is then determined from renormalization group running. 
The CMSSM thus features a very specific 
mass pattern. 

First of all, the assumption of gaugino-mass universality leads to 
$M_1:M_2:M_3 \simeq 1:2:7$ with $M_1\simeq 0.4\,m_{1/2}$ 
at the electroweak scale. Moreover, for not too large scalar soft terms $m_0$ and $A_0$,
$|\mu|^2\gtrsim m_{1/2}^2$. The lightest neutralino is then mostly bino, the second-lightest 
mostly wino, and the heavier ones mostly higgsinos:   
$\nt_1\sim \ti B$, $\nt_2\sim \ti W^3$, $\nt_{3,4}\sim \ti H^0_{1,2}$.
Likewise, $\ch_1\sim \ti W^3$ and $\ch_2\sim \ti H^\pm$. 
Light higgsinos and large gaugino--higgsino mixing occur only for large $m_0$, in the 
so-called focus-point or hyperbolic branch region. 

Furthermore, for the first two generations of squarks one finds
  $M_{\ti U,\ti D}^2\approx m_0^2 + K m_{1/2}^2$,  
  $M_{\ti Q}^2\approx m_0^2 + (K+0.5)\, m_{1/2}^2$
with $K\sim 4.5$ to 6.5. This means that also the squark masses are tightly related to the gluino mass.  
To be more concrete, $m_{\tilde q}\approx m_{\tilde g}$ unless $m_0$ is very large. 
In turn, the gluino mass limit in the CMSSM strongly depends on $m_0$, as can be seen in Fig.~\ref{fig:CMSSMLimit}.  
In fact, for $m_{\tilde q}\gg m_{\tilde g}$, the gluino mass limit goes down to about 750~GeV. 
Furthermore, interpretations within ``Simplified Models'' show \cite{cms:sms} that allowing 
the gluino--LSP mass difference to vary, the limit can become as low as $m_{\tilde g}\gtrsim 400$--500~GeV.

The simplifying assumption of universality at the GUT scale 
makes the model very predictive and a convenient showcase for SUSY phenomenology. 
Indeed, it is interesting to present limits within the CMSSM because it provides 
an easy way to show performances and compare limits or reaches. 
On the other hand, the interpretation of experimental results in the 
$(m_0,m_{1/2})$ plane risks imposing unwarranted constraints on SUSY, as many 
mass patterns and signatures that are possible \emph{a priori} are not covered 
in the CMSSM.

\section{Natural SUSY}

\begin{figure}[t]
\label{fig:NaturalSpectrum}
\begin{center} \includegraphics[width=7cm]{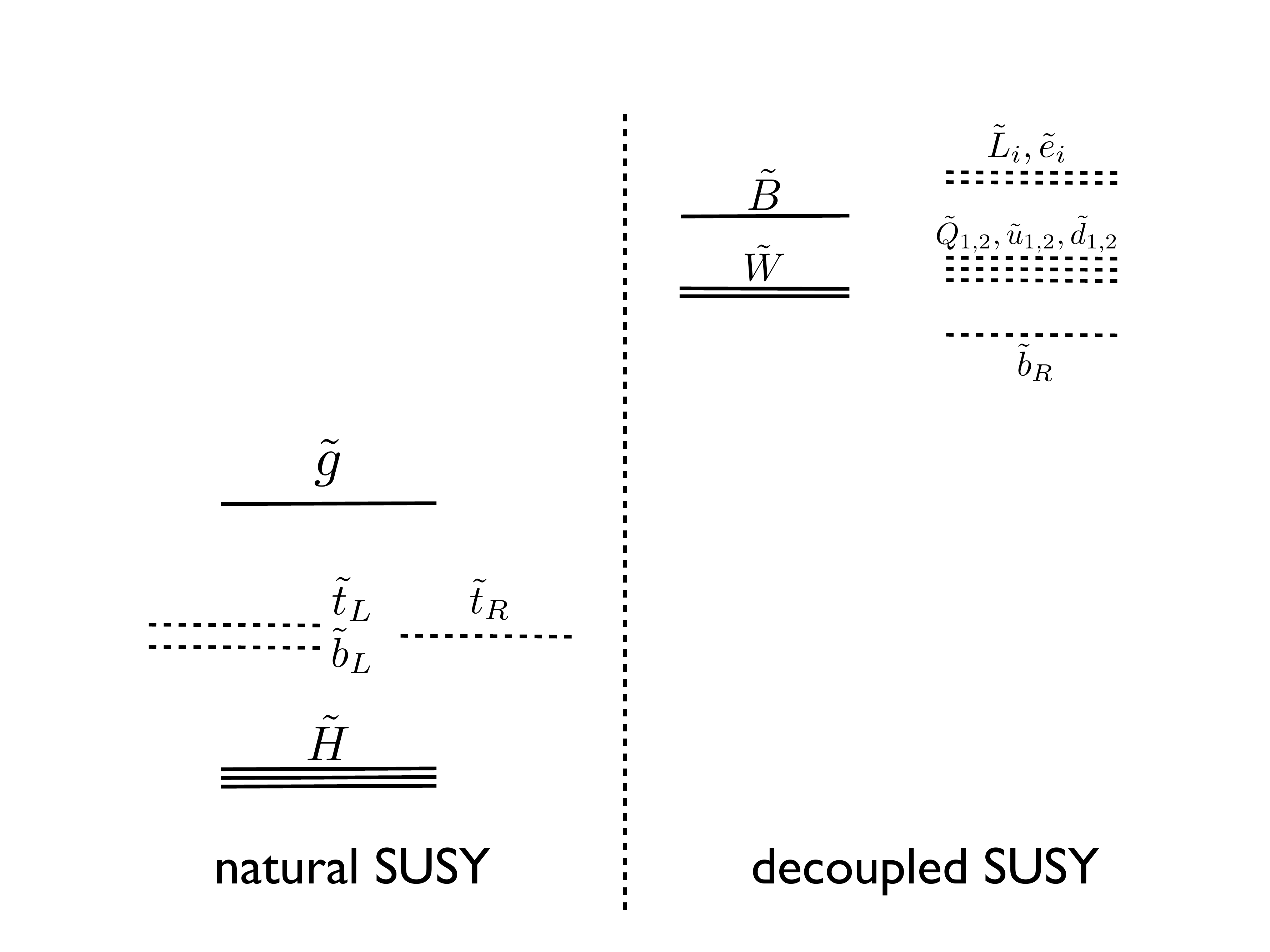} \end{center}
\caption{Natural electroweak symmetry breaking constrains the superpartners on the left to be light.  Meanwhile, the superpartners on the right can be heavy, $M \gg 1$~TeV, without spoiling naturalness. From \cite{Papucci:2011wy}. }
\end{figure}

Natural electroweak symmetry breaking is the leading motivation for why we might expect to discover SUSY particles at the LHC. For the MSSM, the naturalness requirement can be summarized by the tree-level relation 
\be \label{eq:tune}
-\frac{m_Z^2}{2} = |\mu|^2 + m_{H_u}^2.
\ee
Evidently, the contributions to the right-hand side of eq.~(\ref{eq:tune}) must be tuned against each other to achieve electroweak symmetry breaking and the correct $m_Z$.  For SUSY being ``natural'', this tuning should not be too severe. Equation~(\ref{eq:tune}) thus provides  guidance as to which superpartners are required to be light---namely those with the strongest coupling the Higgs. In particular, the higgsinos should not be too heavy because their mass is controlled by $\mu$.  The stop and gluino masses, correcting $m_{H_u}^2$ at one and two-loop order, respectively, also cannot be too heavy.  By SU(2), this also constrains the left sbottom to be light. The masses of the rest of the superpartners, including the squarks of the first two generations, are not important for naturalness and can be much heavier than the present LHC reach. This defines the spectrum of ``natural SUSY'' as depicted in Fig.~\ref{fig:NaturalSpectrum}.

\begin{figure}[t]\centering
\includegraphics[width=126mm]{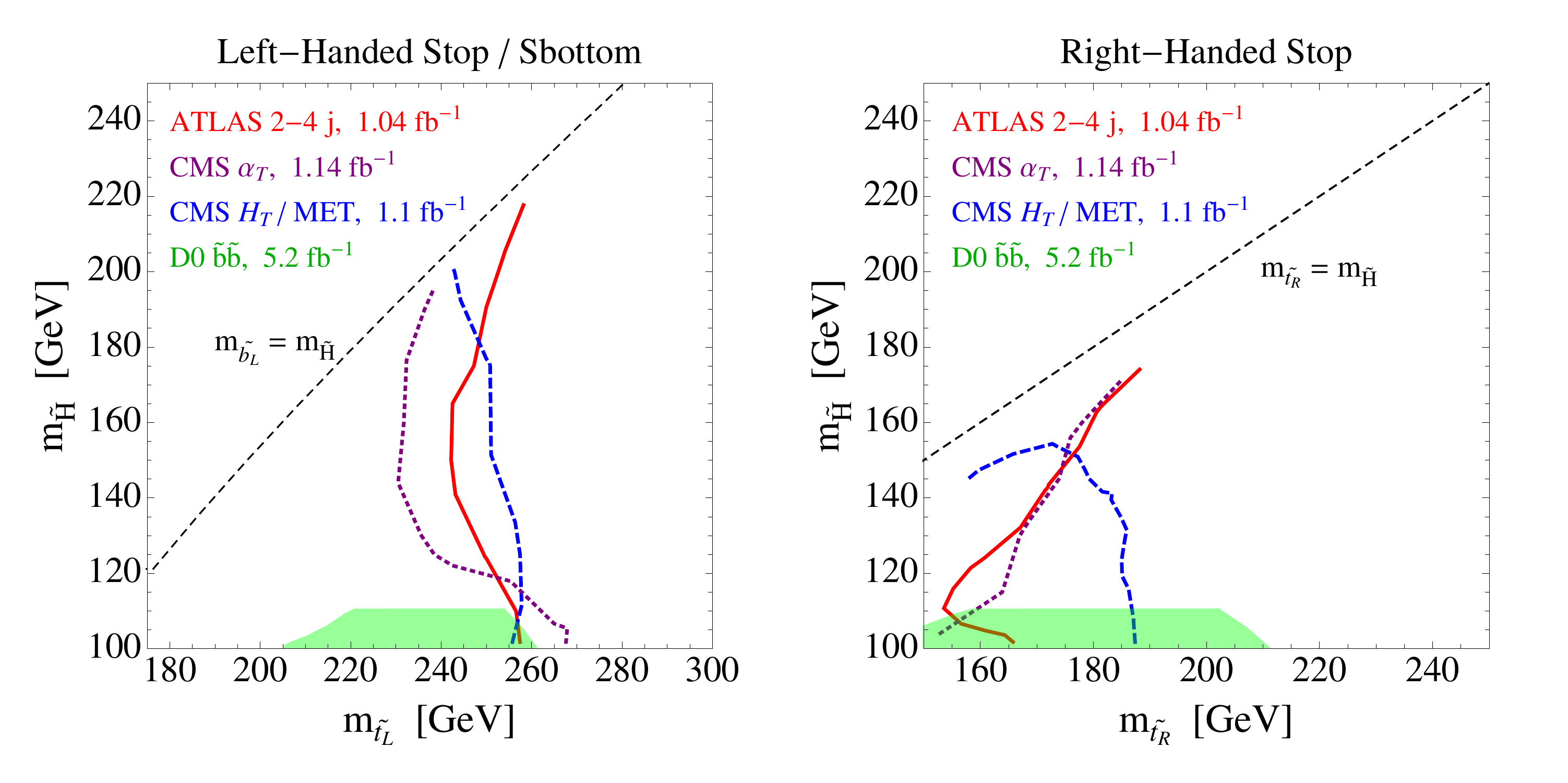}  \vspace*{-2mm}
\caption{ \label{fig:StopHiggsinoLimit}
The LHC limits on the left-handed stop/sbottom ({\it left}) and right-handed stop ({\it right}), with a higgsino LSP.  The axes correspond to the stop pole mass and the higgsino mass.   The strongest limits on this scenario come from searches for jets plus missing energy.   The D0 limit with $5.2\unit{fb}^{-1}$,  shown in green for comparison, only applies for $m_{\tilde \chi^0_1} \lesssim 110$~GeV. From \cite{Papucci:2011wy}. }
\end{figure}

The question to what extend current LHC results constrain (or invalidate) natural SUSY was investigated in \cite{Papucci:2011wy}. It turned out that the strongest limits  come from searches for jets plus missing energy. The result of the analysis is shown in Fig.~\ref{fig:StopHiggsinoLimit}. There is a stronger limit on the left-handed stop (left plot) than the right-handed stop (right plot), because of the additional presence of a sbottom, in the left-handed case, leading to an overall larger production cross-section than for the right-handed stop.  In both cases the limits are set by both stops and bottoms decaying to b-jets and chargino or neutralino respectively.

The limits based on 1\fbi\ at 7 TeV are roughly $m_{\tilde t_L}\gtrsim 260$~GeV and $m_{\tilde t_R}\gtrsim 190$~GeV. 
We conclude that LHC searches do not yet significantly constrain natural SUSY.

\clearpage

\section{Phenomenological MSSM}

The question to ask next is what current LHC data really tell us, and do not tell us, about the SUSY in general. Of course the $\gtrsim100$ parameters of the general MSSM are too many to scan over, not to speak  of extensions of the MSSM. With a few well-motivated assumptions we can, however, greatly reduce the dimensionality of the problem: assuming that 
R-parity is conserved, there are no new CP phases, the sfermion mass matrices and trilinear couplings are flavor-diagonal, 
the first two generations of sfermions are degenerate and their trilinear couplings are negligible, we arrive at the so-called \emph{phenomenological MSSM} (pMSSM)~\cite{Djouadi:1998di}. (We also assume that the lightest neutralino is the LSP.)
The pMSSM has 19 free parameters defined at the SUSY scale,  
$M_{\rm SUSY}\equiv \sqrt{m_{\tilde t_1}m_{\tilde t_2}}$:
\begin{itemize}
   \item the gaugino mass parameters $M_1$, $M_2$, and $M_3$; 
   \item the ratio of the Higgs VEVs $\tan\beta=v_2/v_1$;
   \item the higgsino mass parameter $\mu$ and 
            the pseudo-scalar Higgs mass $m_A$;
    \item 10 sfermion mass parameters $m_{\tilde{F}}$, where 
         $\tilde{F} = \tilde{Q}_1, \tilde{U}_1, \tilde{D}_1, 
                      \tilde{L}_1, \tilde{E}_1, 
                      \tilde{Q}_3, \tilde{U}_3, \tilde{D}_3, 
                      \tilde{L}_3, \tilde{E}_3$\\ 
(imposing $m_{\tilde{Q}_1}\equiv m_{\tilde{Q}_2}$, 
           $m_{\tilde{L}_1}\equiv m_{\tilde{L}_2}$, etc.), and          
   \item 3 trilinear couplings $A_t$, $A_b$ and $A_\tau$\,,               
\end{itemize}
in addition to the SM parameters, and is thus independent of any SUSY breaking scheme. 
Indeed it is important to note that~\cite{Berger:2008cq}  
``the pMSSM leads to a much broader set of predictions for the properties 
of the SUSY partners as well as for a number of experimental observables than 
those found in any of the conventional SUSY breaking scenarios such as mSUGRA [CMSSM]. 
This set of models can easily lead to atypical expectations for SUSY signals 
at the LHC.'' 
 
In \cite{Sekmen:2011cz}, taking a Bayesian approach, we interpreted  the results of SUSY 
searches published by the CMS collaboration based on the first $\sim$1~fb$^{-1}$ of data 
at 7~TeV within the pMSSM, thus deriving constraints 
on the SUSY particles with as few simplifying assumptions as possible. 
To be concrete, we used the results from three independent CMS analyses: the 
$\alpha_T$~hadronic~\cite{Collaboration:2011zy}, 
the same-sign (SS) dilepton~\cite{CMS:SS} and 
the opposite-sign (OS) dilepton~\cite{CMS:OS} analyses. 

The initial sampling was done by means of a Markov Chain Monte Carlo (MCMC), for which we used a number of ``preLHC'' constraints such as $BR(b \rightarrow s\gamma)$, $BR(B_s \rightarrow \mu \mu)$, $\Delta a_\mu$, and Higgs mass limits, see \cite{Sekmen:2011cz} for details. (Note that we not to impose any constraint on $\Omega h^2$.) From  $1.5\times10^7$ collected Markov-chain points we then drew a subset of $5\times10^5$ points, for each of 
which we generated 10K events using {\tt PYTHIA6}~\cite{Sjostrand:2006za}. 
The response of the CMS detector was mimicked using {\tt Delphes}~\cite{Ovyn:2009tx}. The simulated event count for the $\alpha_T$~hadronic, SS and OS dilepton analyses (with selection criteria as described in \cite{Collaboration:2011zy,CMS:SS,CMS:OS}) was then compared to the observed event counts and background estimates from the official CMS results~\cite{Collaboration:2011zy,CMS:SS,CMS:OS}. The final posterior probability is then approximated by
weighting each pMSSM point by the ``CMS likelihood'' from each of the three analyses. 

\begin{figure}[t]
   \centering
   \includegraphics[width=3.6cm]{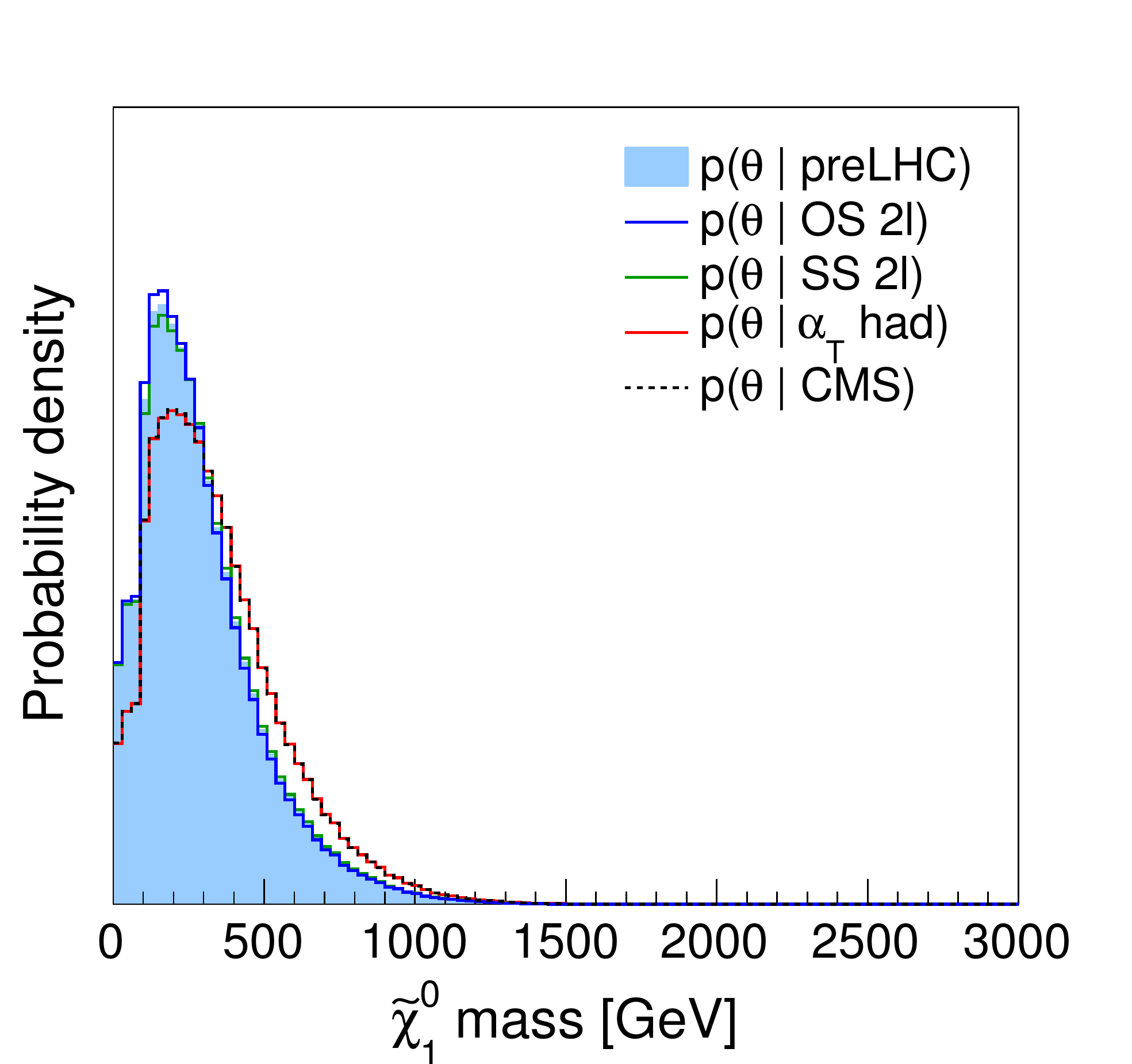} 
   \includegraphics[width=3.6cm]{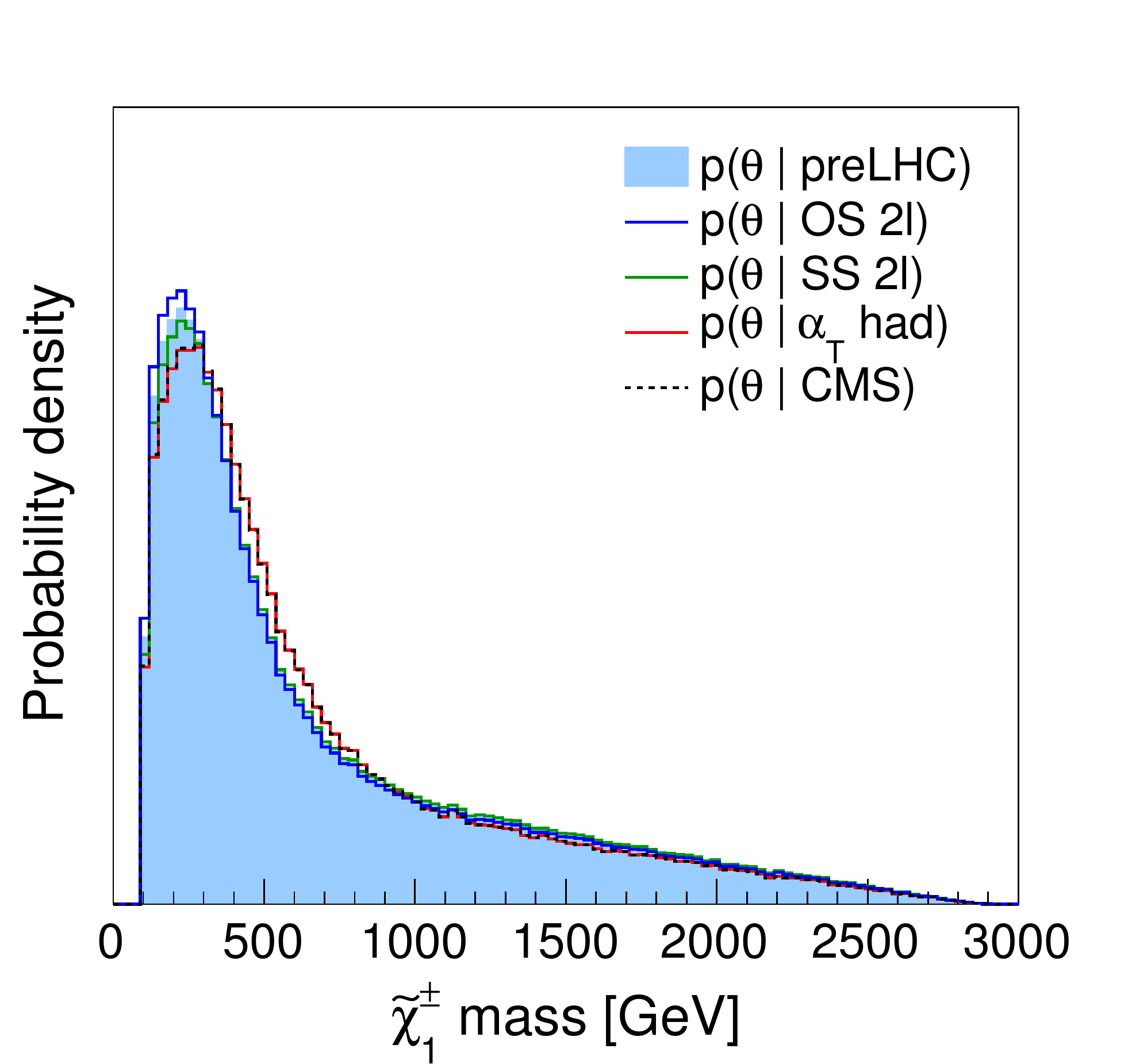} 
   \includegraphics[width=3.6cm]{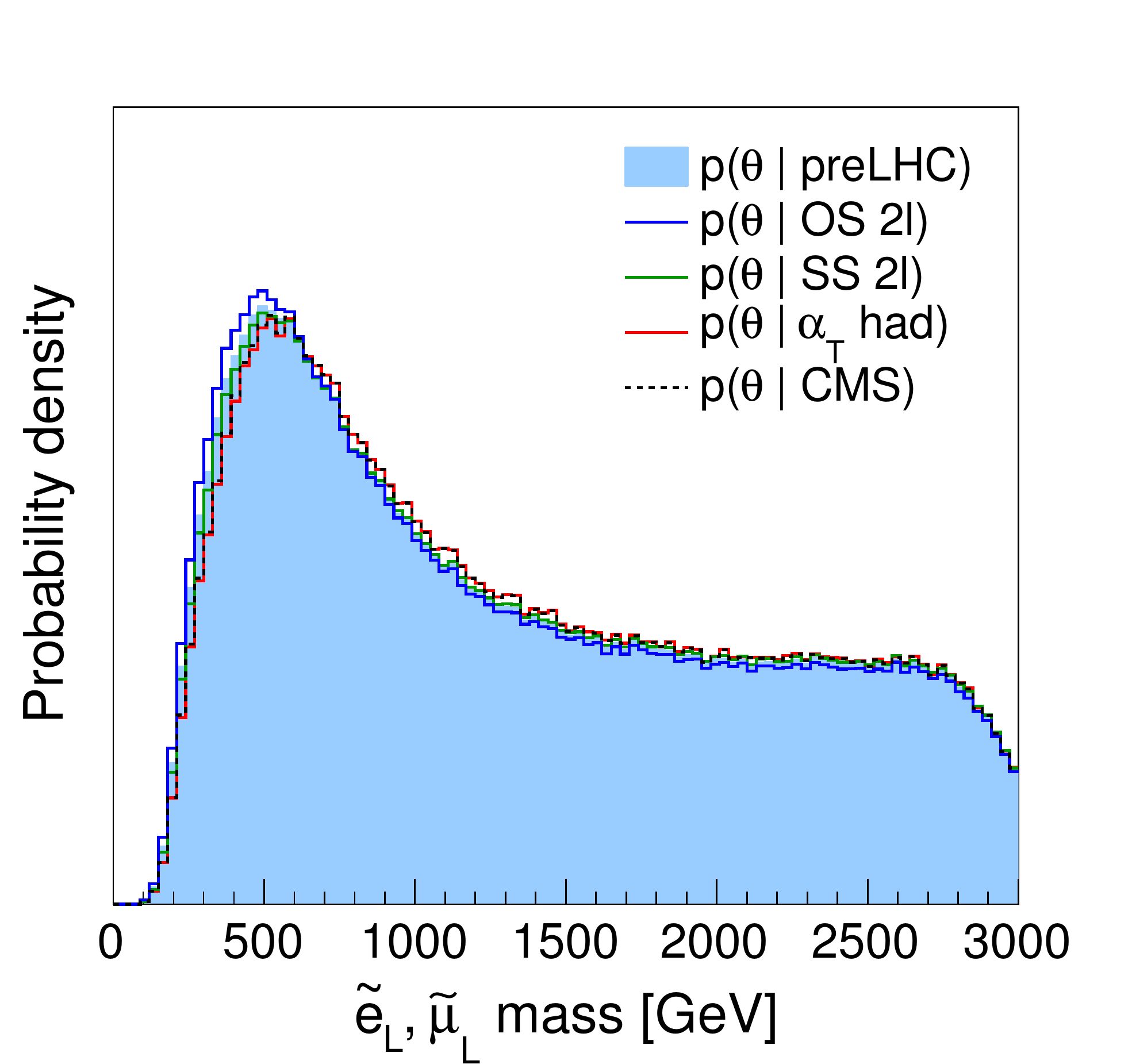} 
   \includegraphics[width=3.6cm]{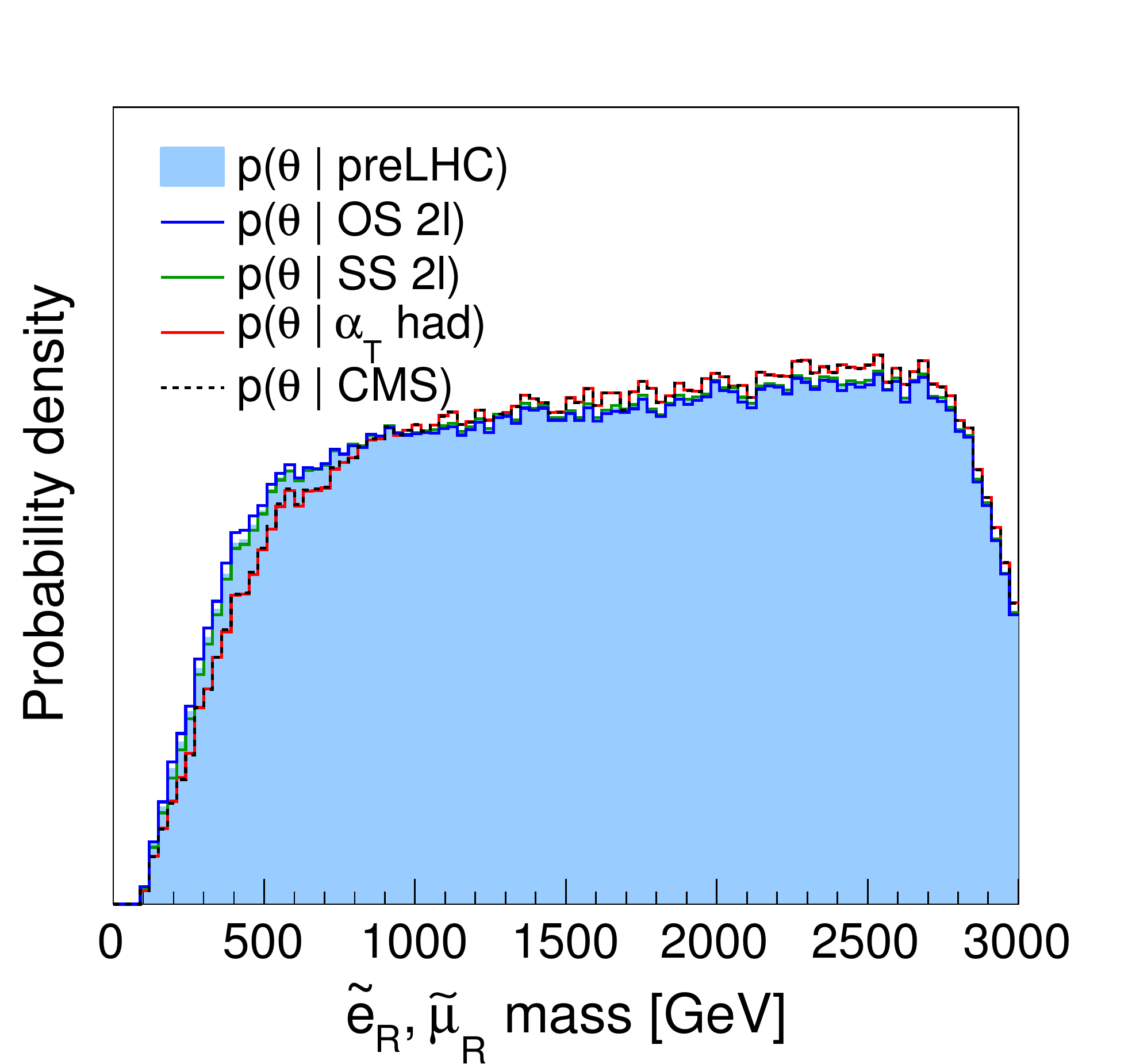} 
   \includegraphics[width=3.6cm]{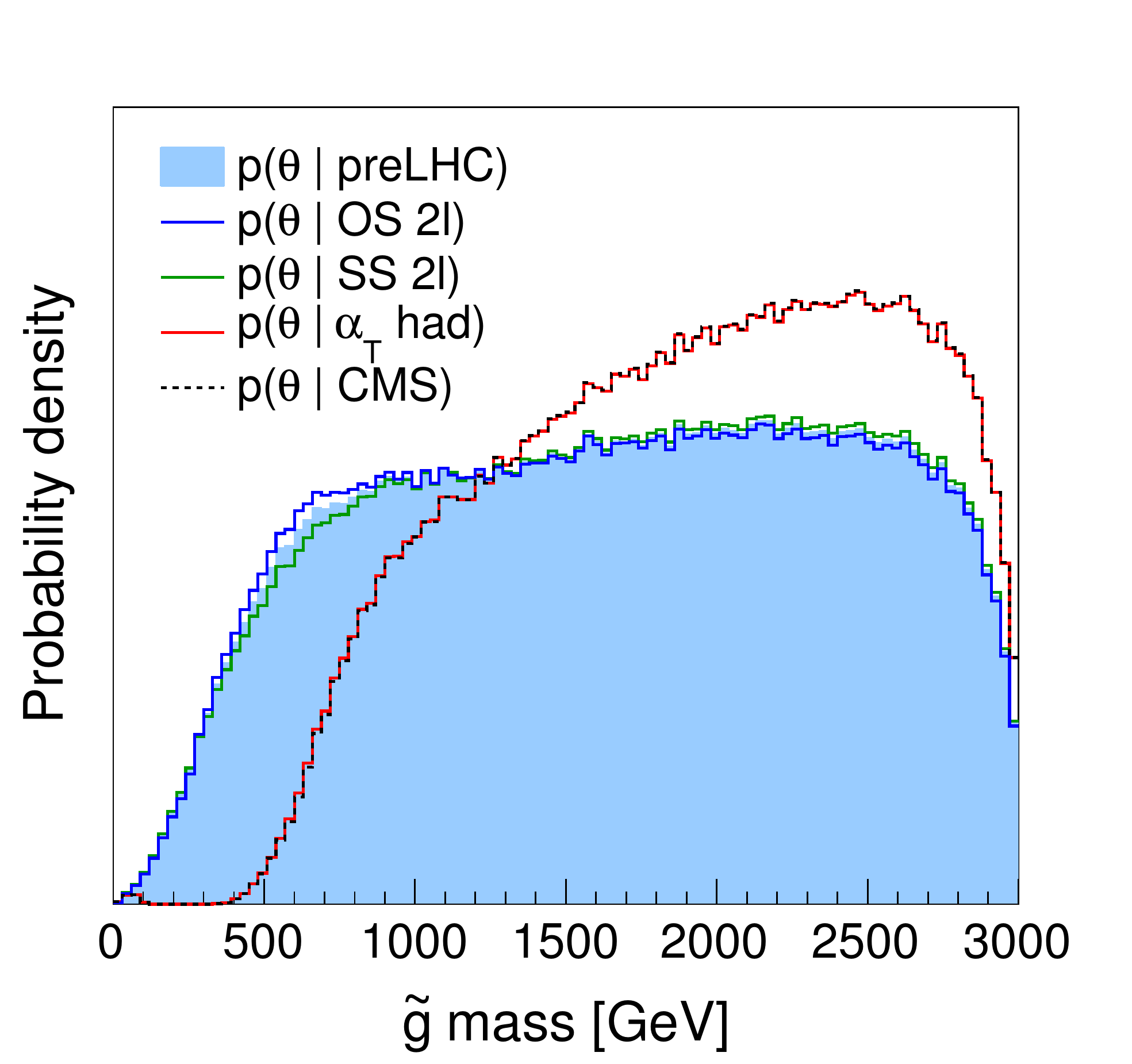} 
   \includegraphics[width=3.6cm]{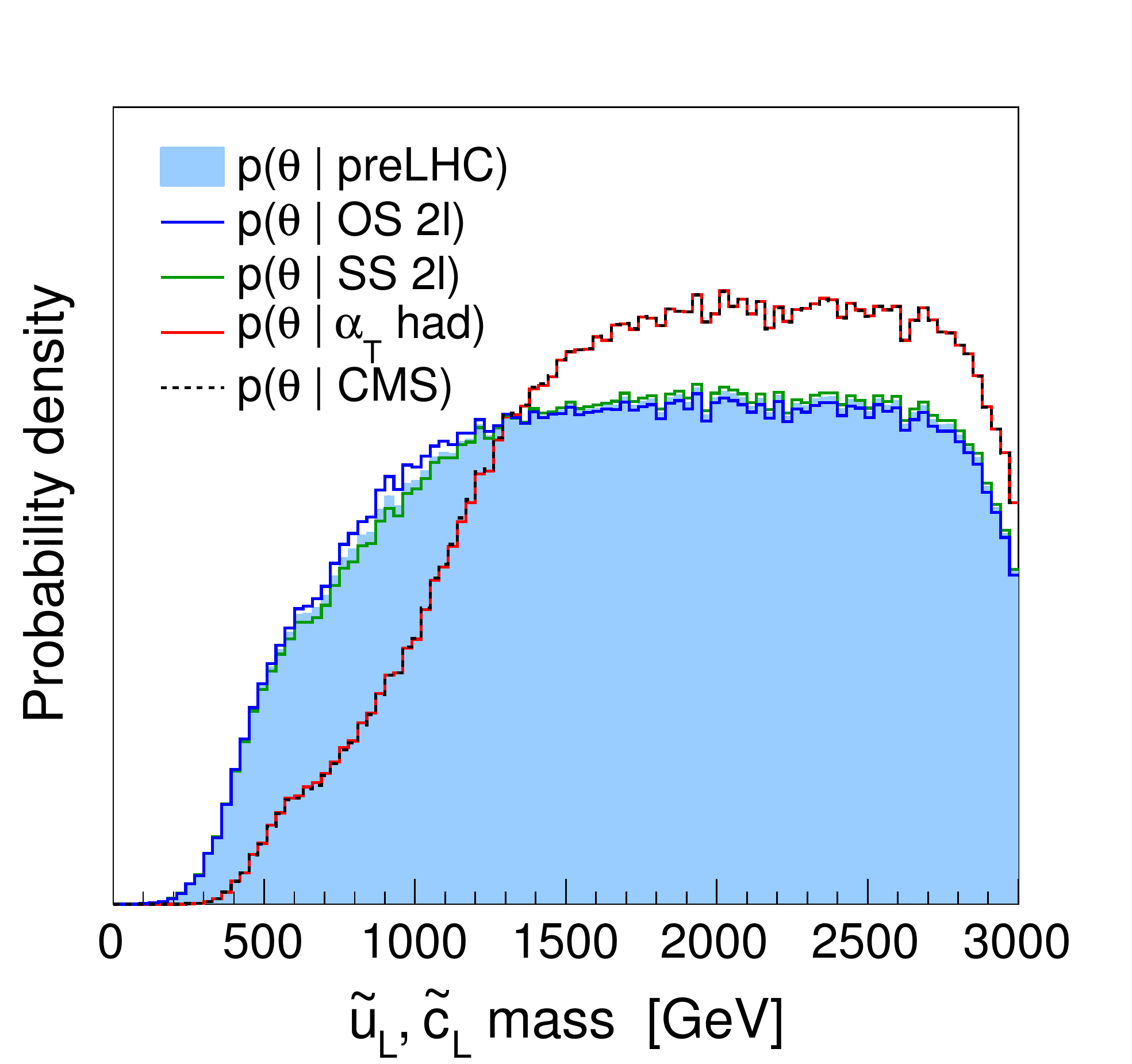} 
   \includegraphics[width=3.6cm]{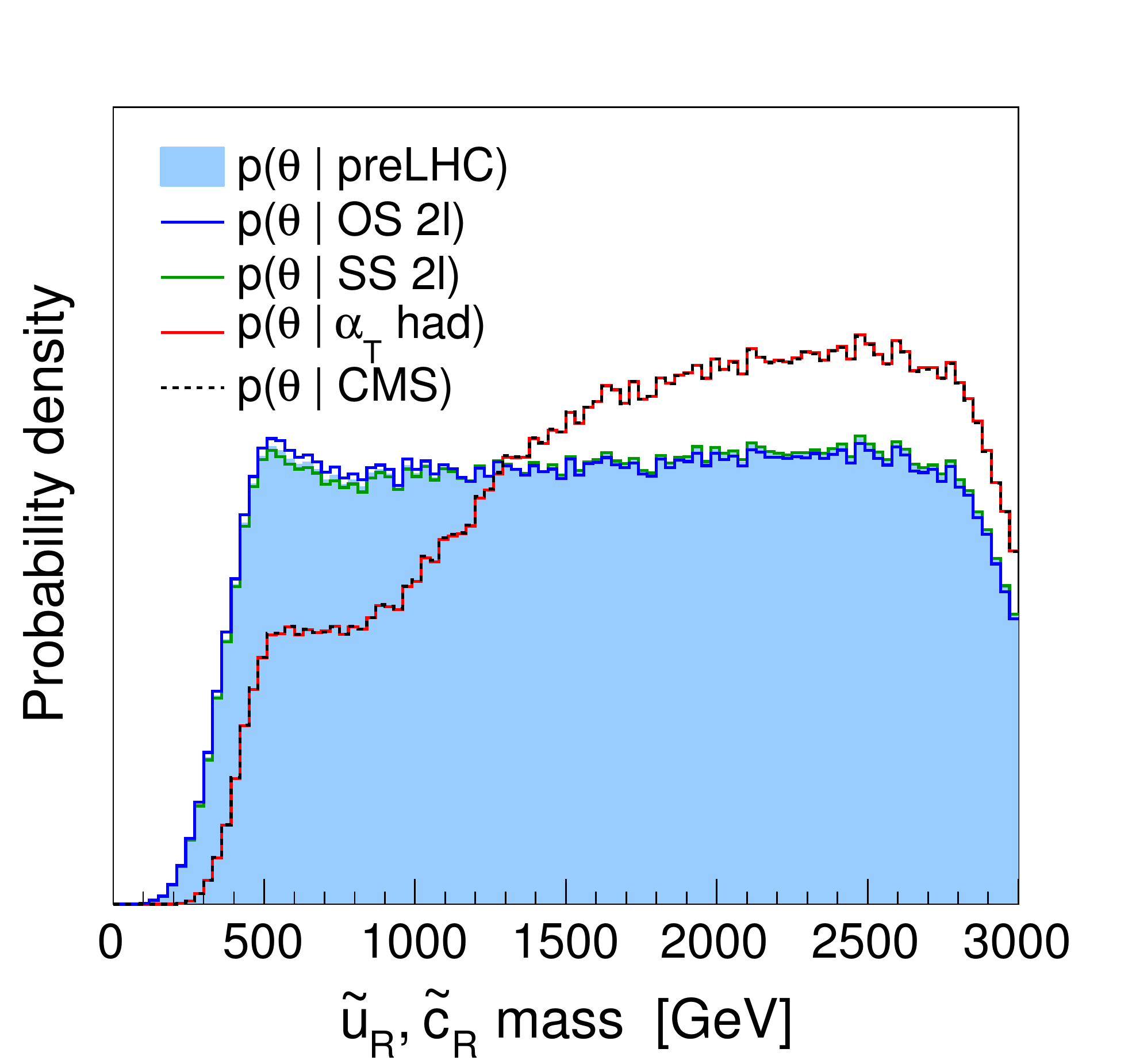} 
   \includegraphics[width=3.6cm]{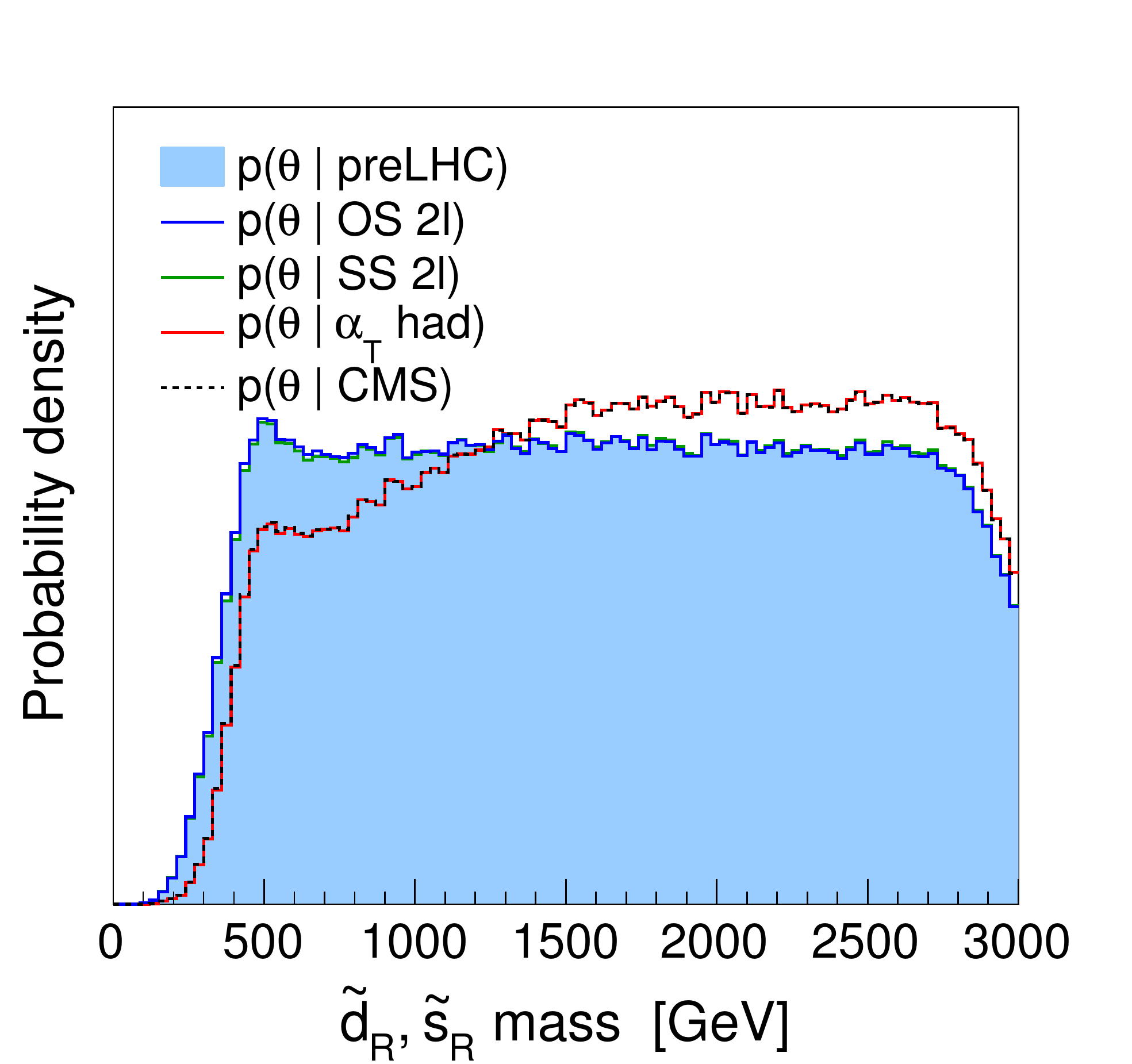} 
   \includegraphics[width=3.6cm]{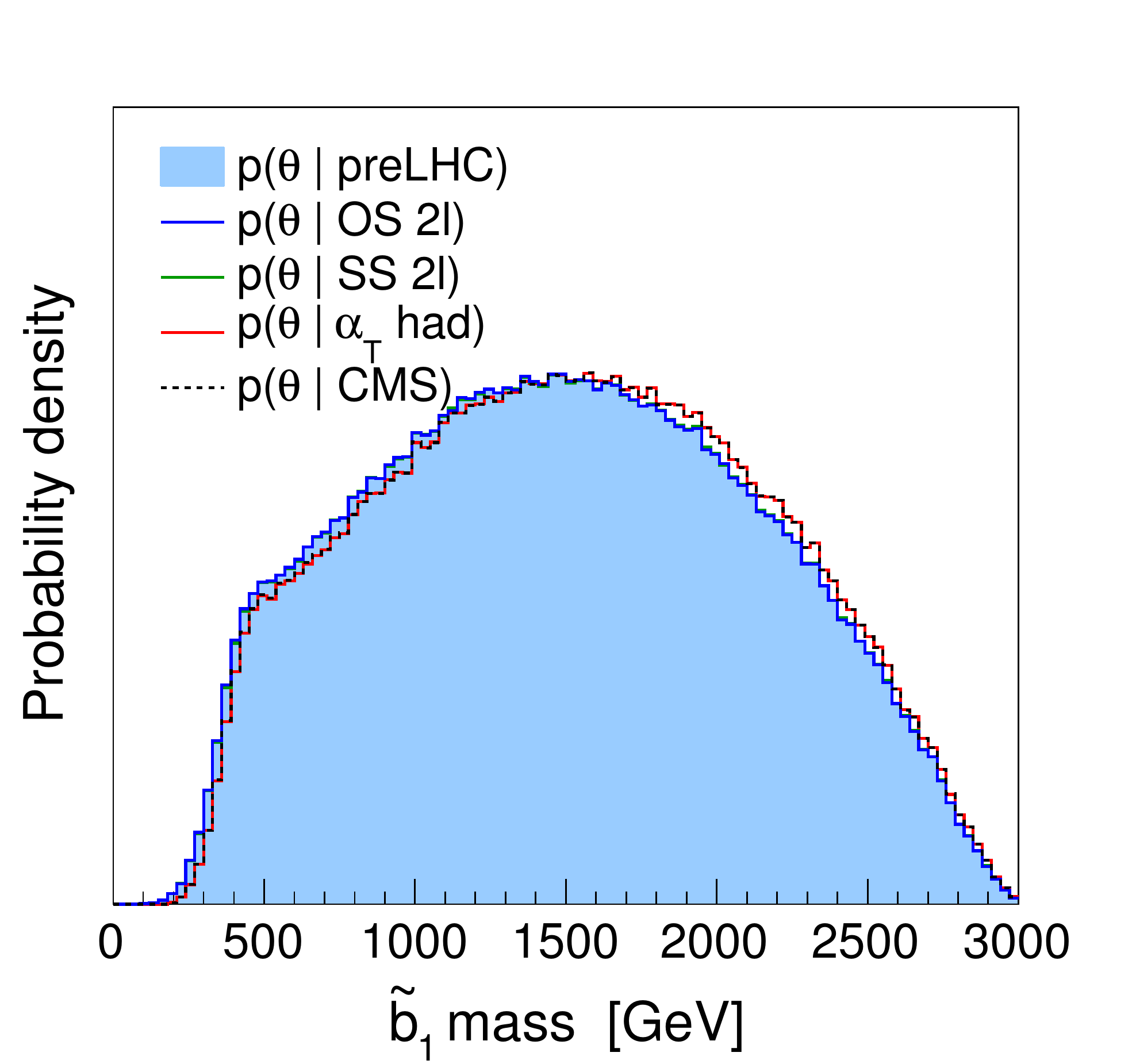} 
   \includegraphics[width=3.6cm]{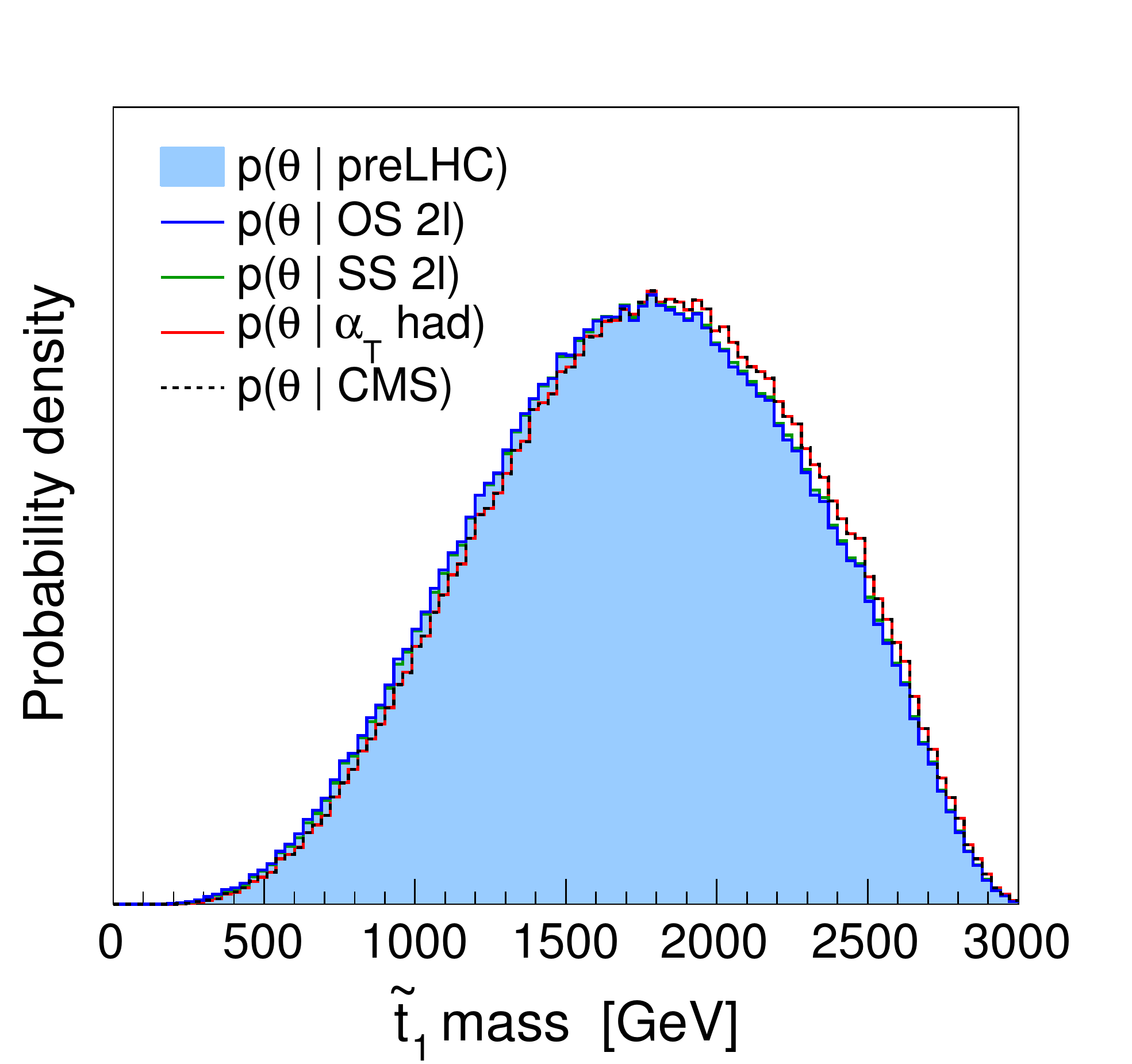} 
   \includegraphics[width=3.6cm]{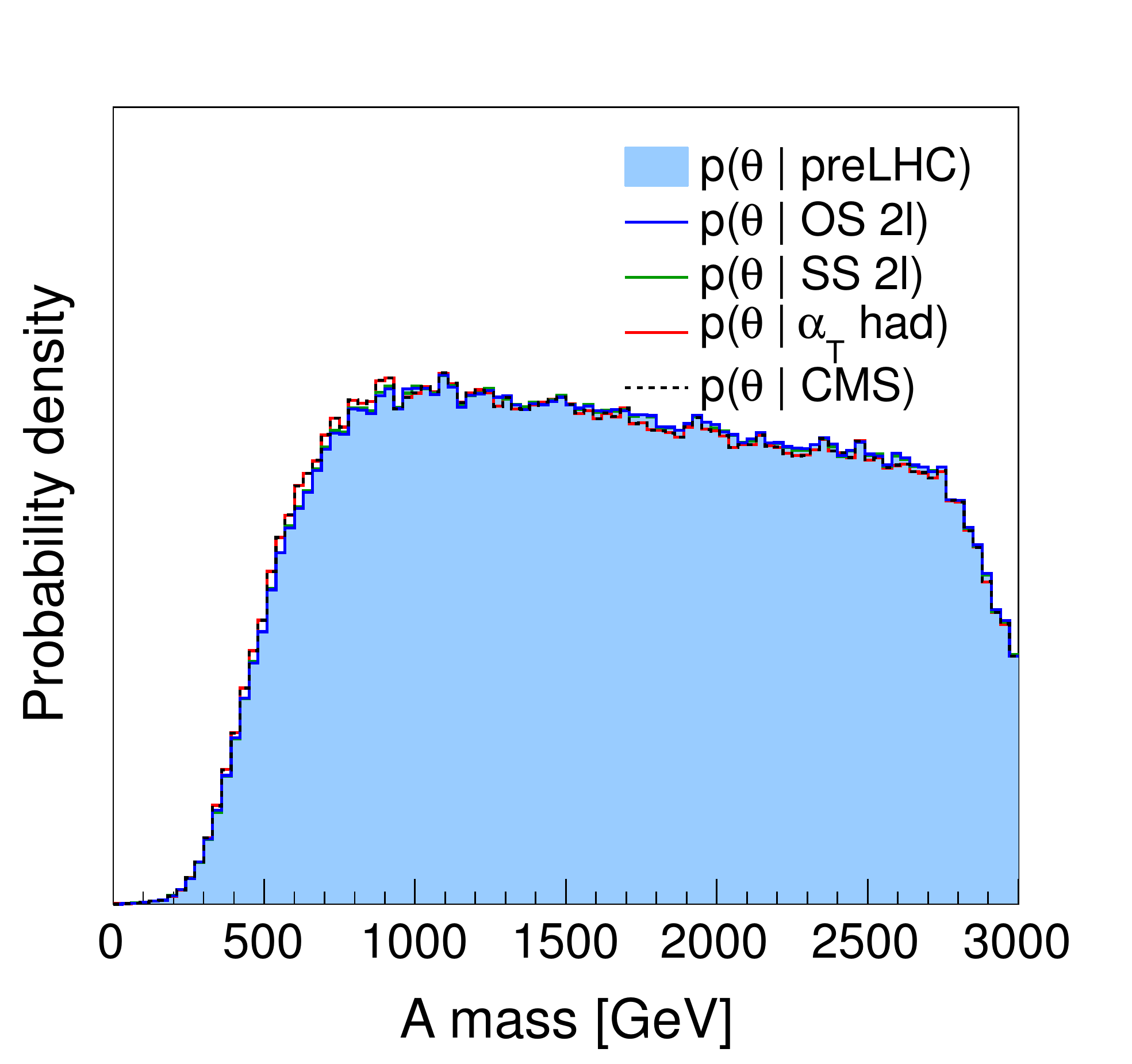} 
   \includegraphics[width=3.6cm]{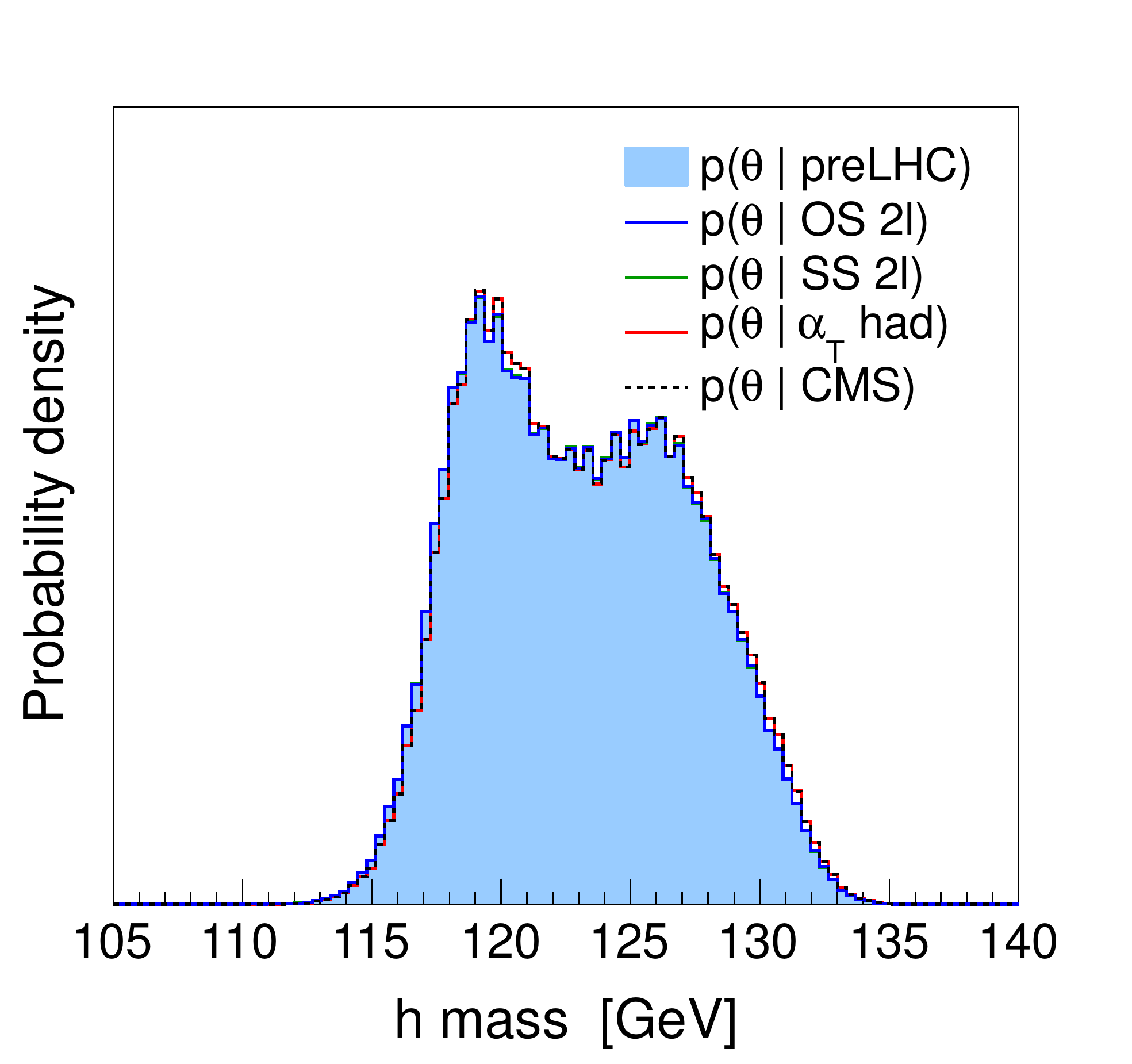} 
   \caption{Marginalized 1D posterior densities of sparticle and Higgs masses.}
   \label{fig:dist-1d-masses}
\end{figure}
\begin{figure}[t]
   \centering
   \includegraphics[width=3.6cm]{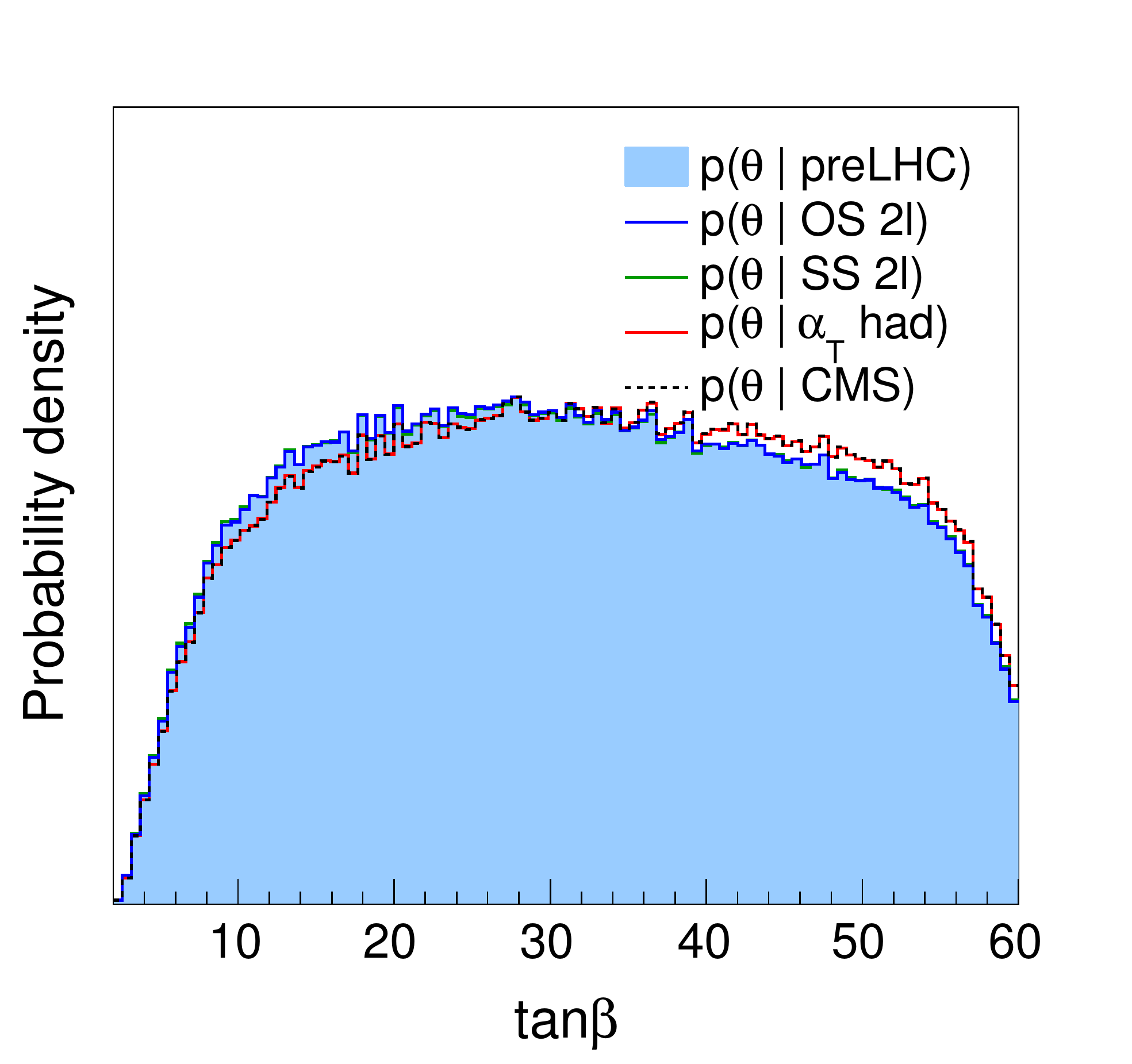} 
   \includegraphics[width=3.6cm]{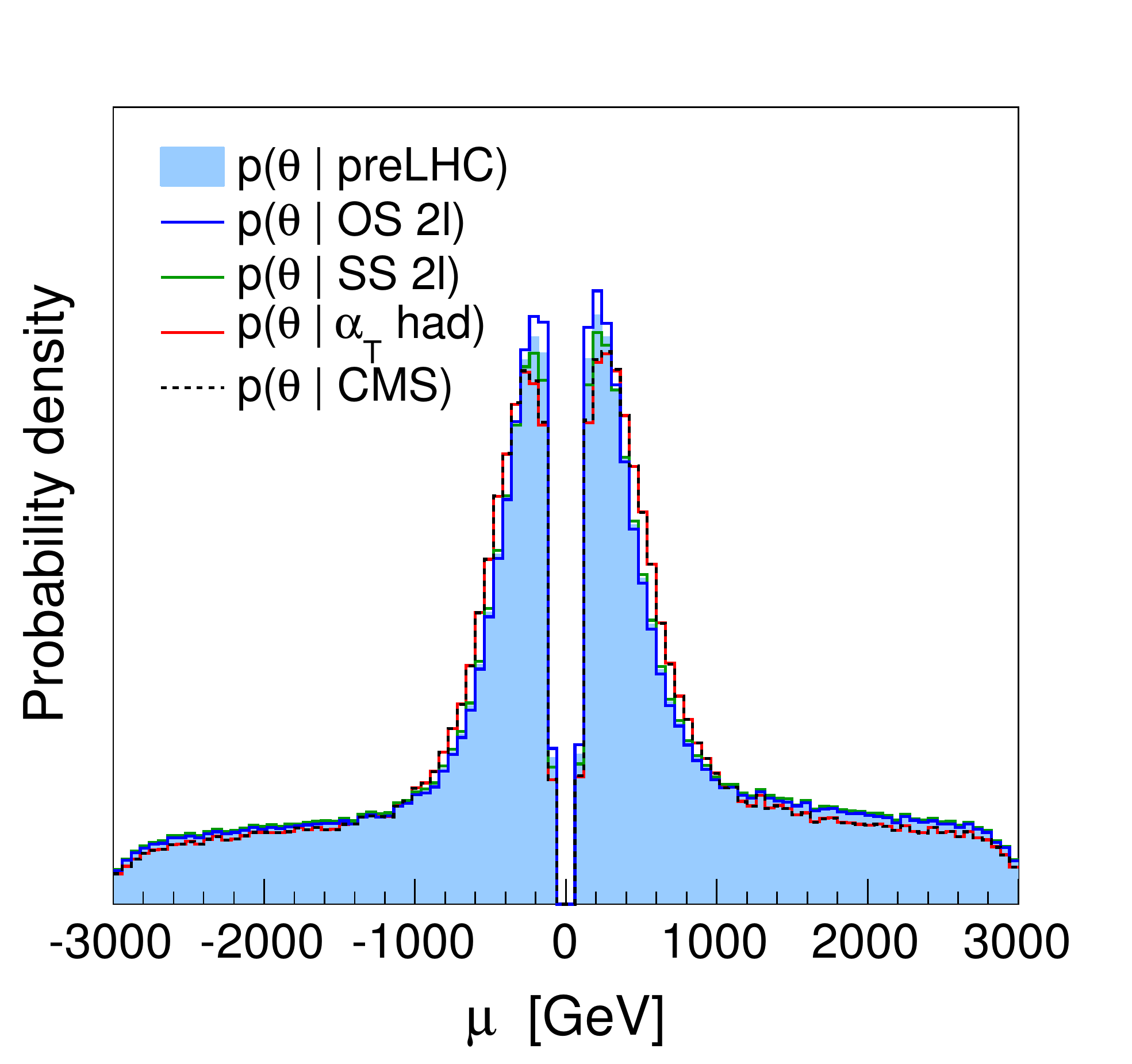} 
   \includegraphics[width=3.6cm]{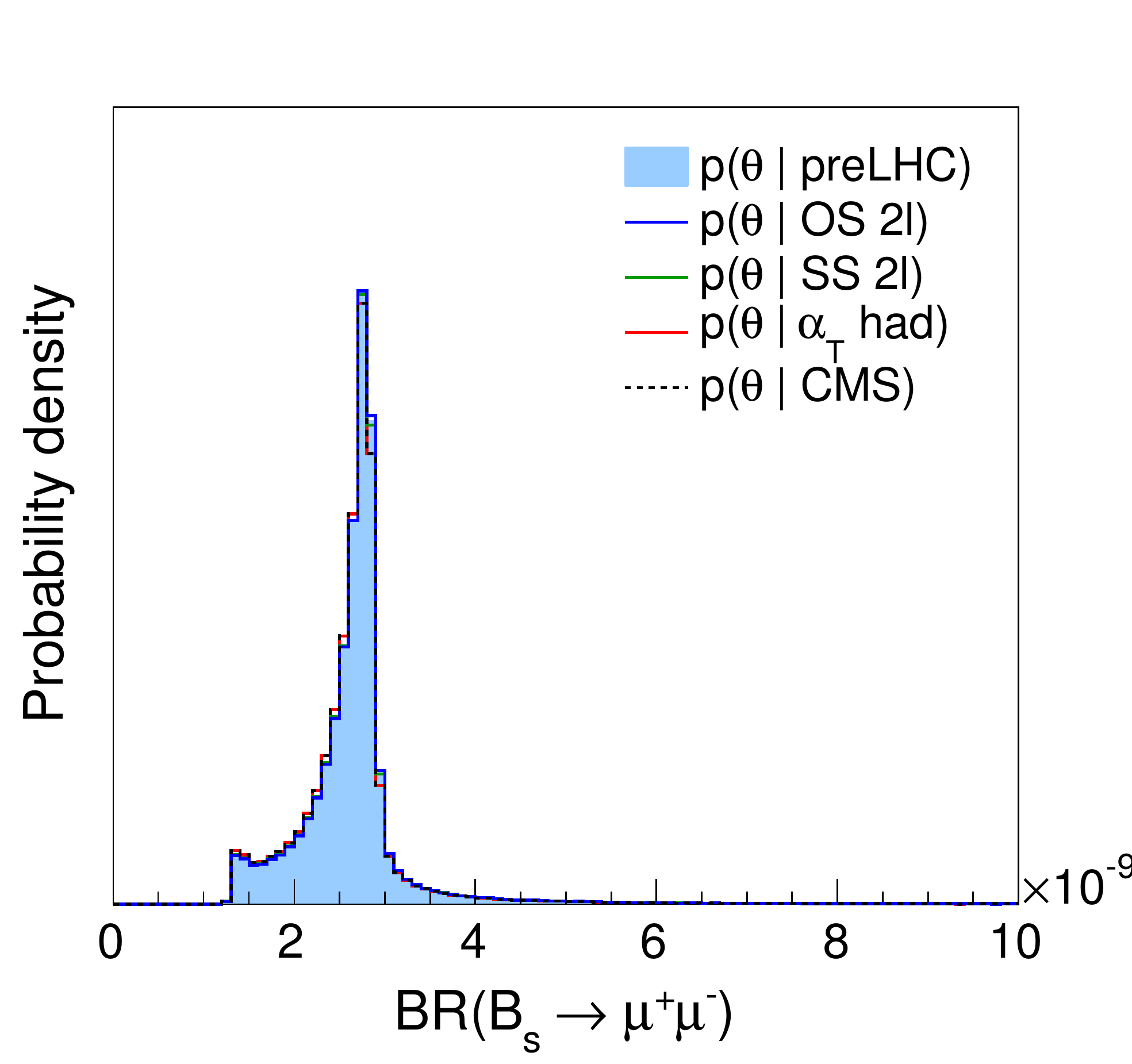} 
   \includegraphics[width=3.6cm]{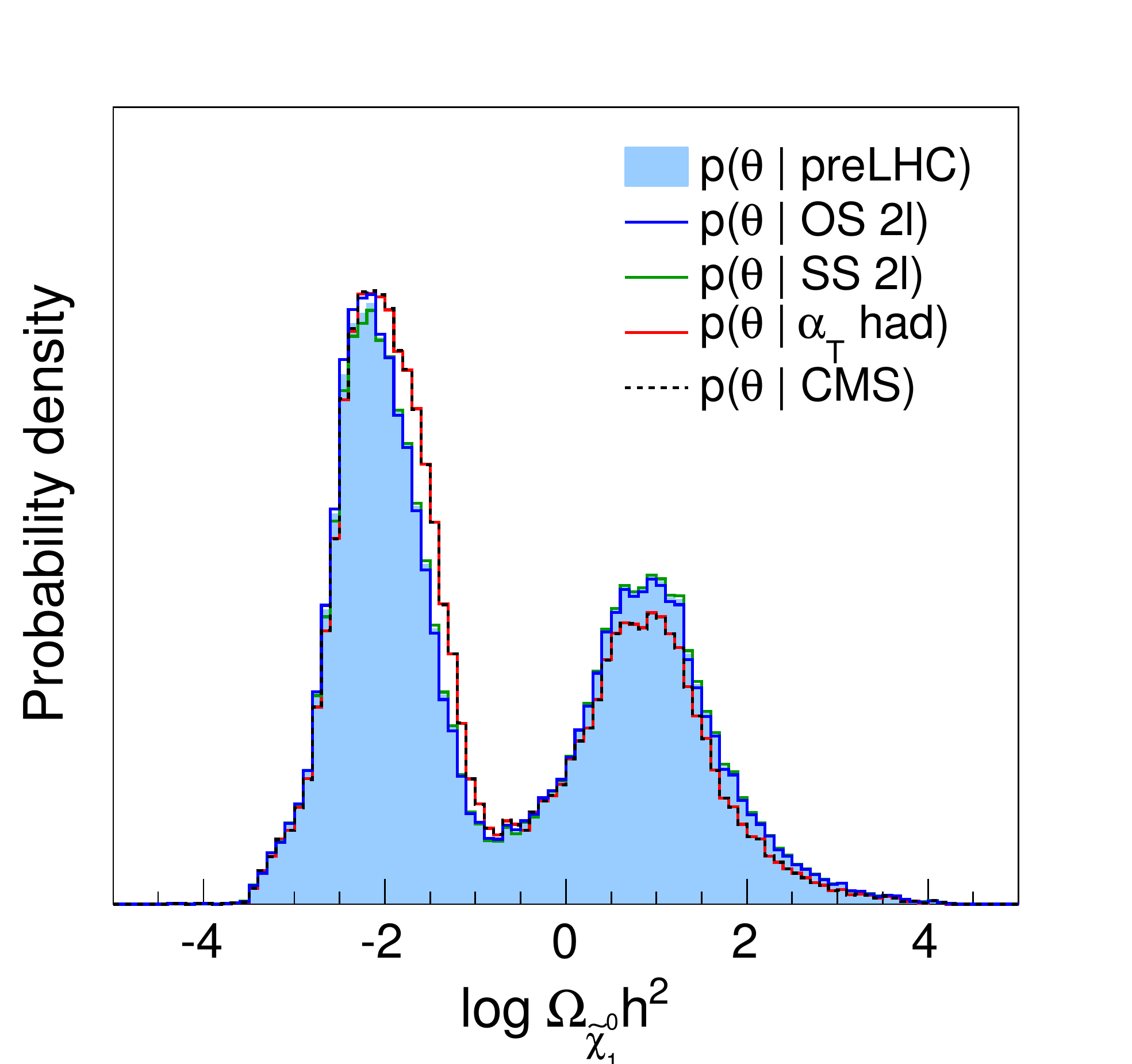} 
   \caption{Marginalized 1D posterior densities of $\tan\beta$, $\mu$, $BR(B_s\to \mu^+\mu^-)$ and neutralino relic density $\Omega h^2$.}
   \label{fig:dist-1d-more}
\end{figure}

Figure~\ref{fig:dist-1d-masses} shows marginalized 1-dimensional (1D) posterior probability density functions  of various sparticle and Higgs masses resulting from this analysis. The light blue histograms represent the preLHC probability densities. (The $\tilde\chi^\pm_1$ and $\tilde e_L/\tilde \mu_L$ are bound to be light by the $\Delta a_\mu$ constraint.) 
The blue, green and red lines show, respectively, the effects of the OS di-lepton, SS di-lepton and  
$\alpha_T$ hadronic CMS analyses. The dashed black lines show the final 
posterior densities after inclusion of the results of all three analyses. 
It is evident that with current LHC data-sets, the di-lepton analyses have very little effect on the 
posterior densities, while the $\alpha_T$ hadronic 
analysis pushes the gluino and $1^{st}/2^{nd}$-generation squark masses towards higher values. 
(What is relevant here is any change in the onset of the posterior distribution, not the high-mass behavior.)
We also note the slight effect on the $\tilde\chi^0_1$ LSP mass. The masses of other sparticles, 
including charginos, sleptons and $3^{rd}$-generation squarks,  are basically unaffected by the 
current LHC results. This contrasts with the CMSSM case, in which all these masses are correlated 
through their dependence on $m_{1/2}$ and $m_0$. 

Finally, we see that the Higgs mass distributions, 
including that of $m_h$, remain unaffected by current SUSY searches. 
It is interesting to note that the Higgs mass window of $m_h=123\mbox{--}128$~GeV has 27.4\% probability in our analysis;  this decreases only marginally to 26.4\% when requiring $\Omega h^2<0.136$. 
Likewise, it is interesting that the SUSY mass distributions in Fig.~\ref{fig:dist-1d-masses} remain unaffected by requiring $m_h=123\mbox{--}128$~GeV; the only impact is in fact on $A_t$ and the stop mixing parameter $X_t/M_S$~\cite{Brooijmans:2012yi}. 
For a discussion of the implications of a 125~GeV Higgs for SUSY, let me refer to the talk by Nazila Mahmoudi~\cite{nazila125}.

The probability of finding $\Omega h^2<0.136$ is 55\%, while $0.094<\Omega h^2<0.136$ has 1\% probability, see the right-most plot in Fig.~\ref{fig:dist-1d-more}. The other plots in this figure show the 1D posterior distributions of $\tan\beta$, $\mu$, $BR(b\to s\gamma)$ and $BR(B_s\to \mu^+\mu^-)$.  Note that the $\tan\beta$ distribution is almost flat for $\tan\beta\approx 10\mbox{--}50$, and that we observe an only marginal preference for $\mu>0$ with $p(\mu>0)\approx0.53$, both pre- and post-LHC results. 

\begin{figure}[t]
   \centering
   \includegraphics[width=3.6cm]{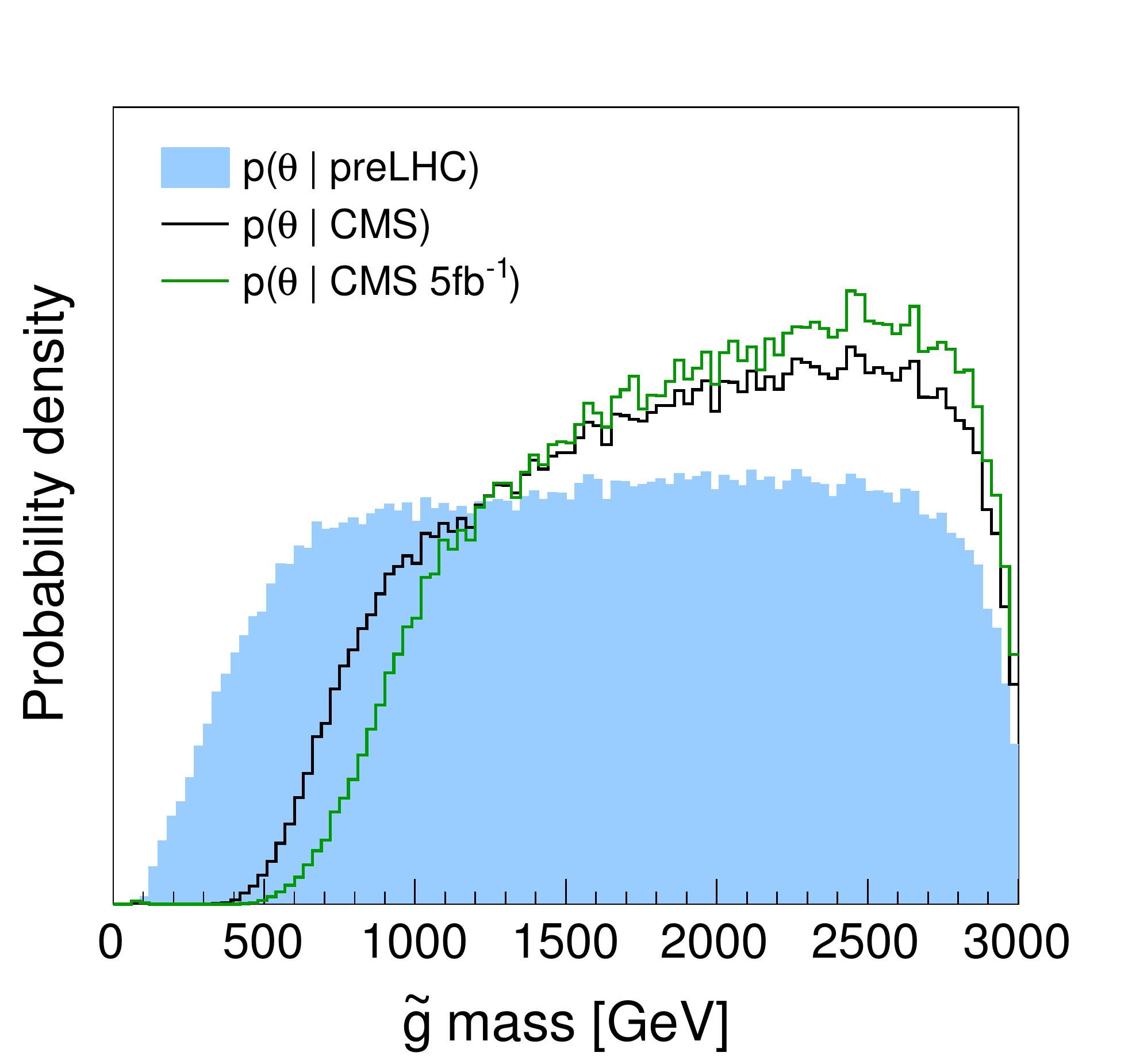} 
   \includegraphics[width=3.6cm]{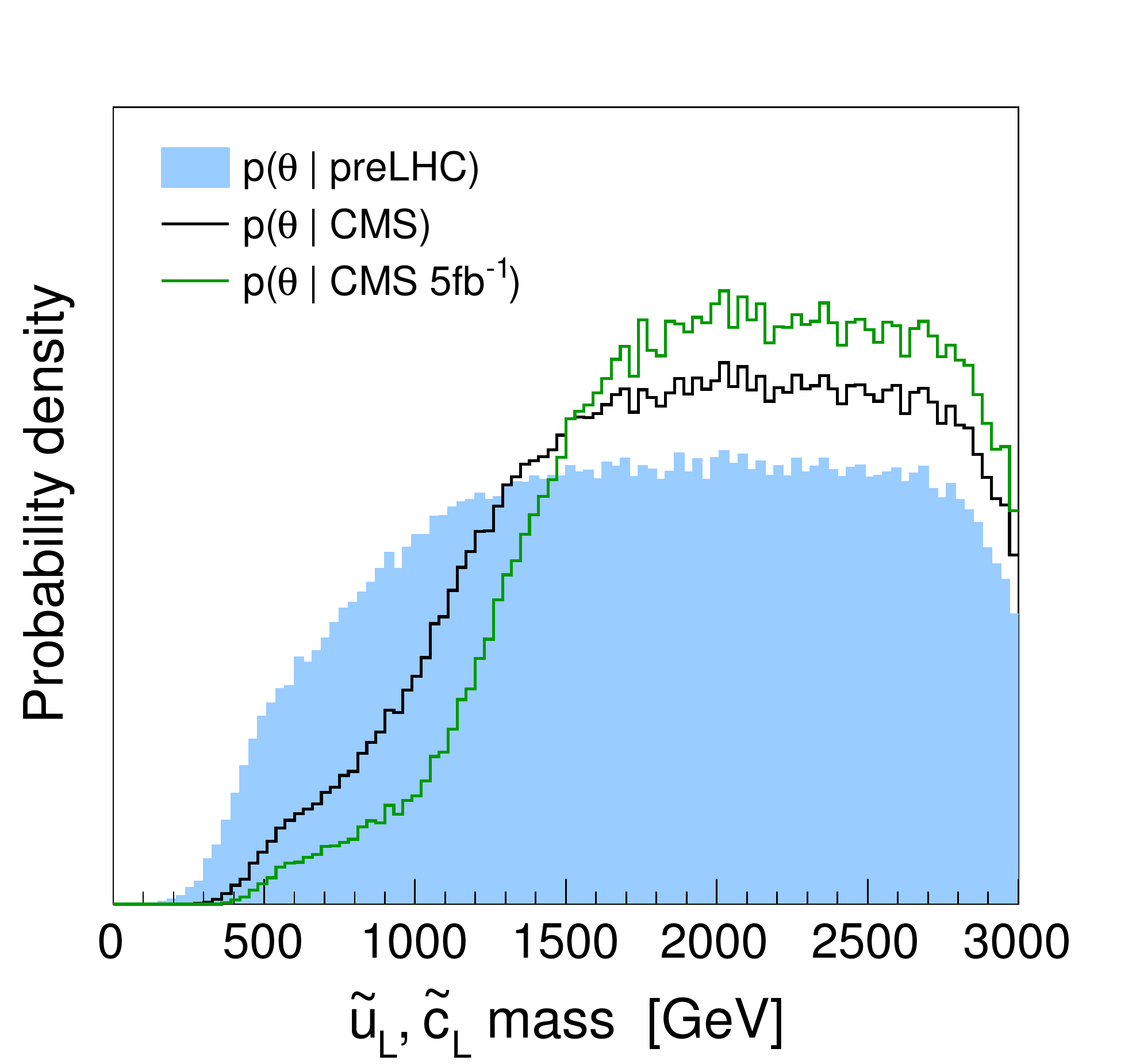} 
   \includegraphics[width=3.6cm]{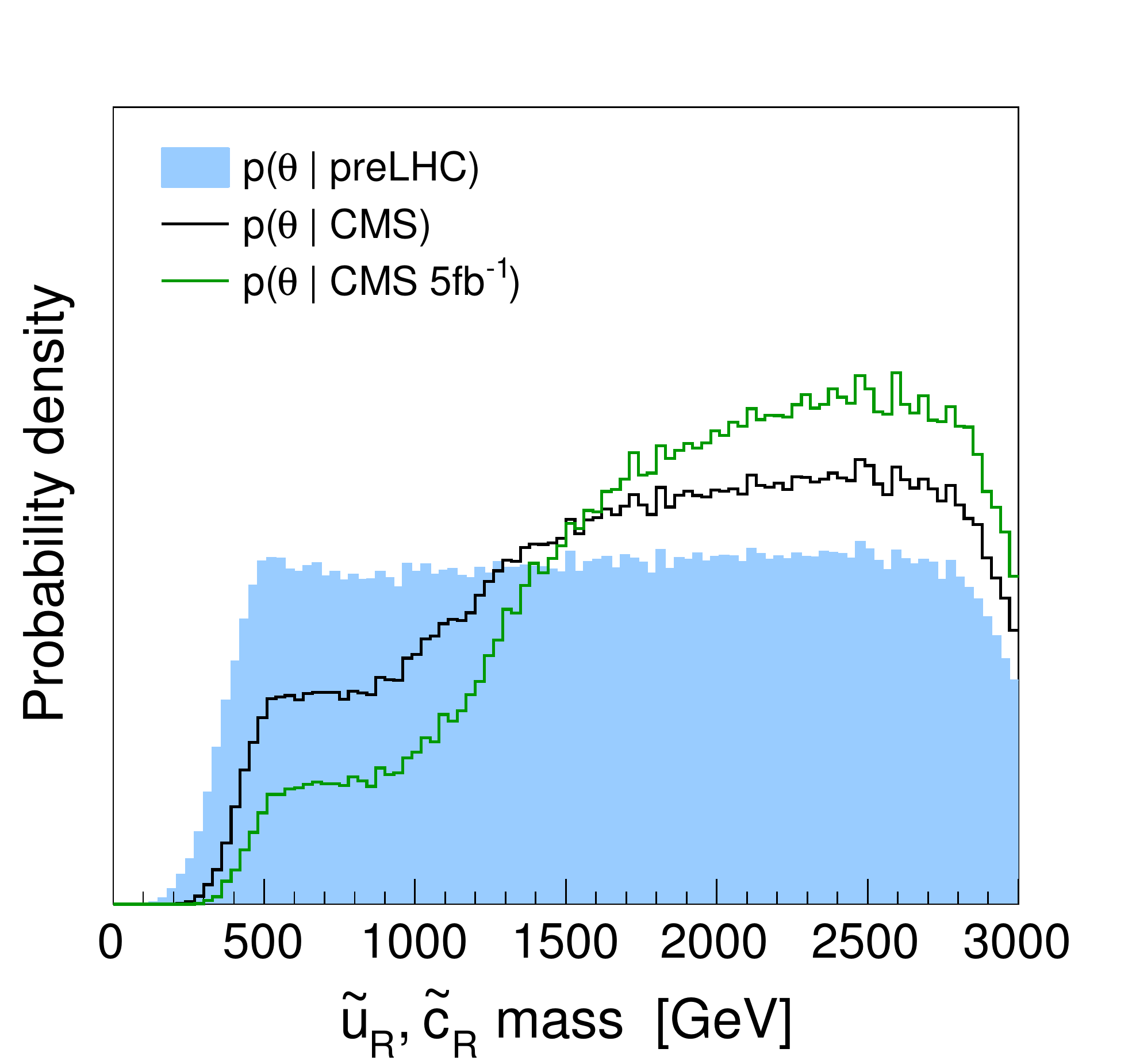} \\
   \includegraphics[width=3.6cm]{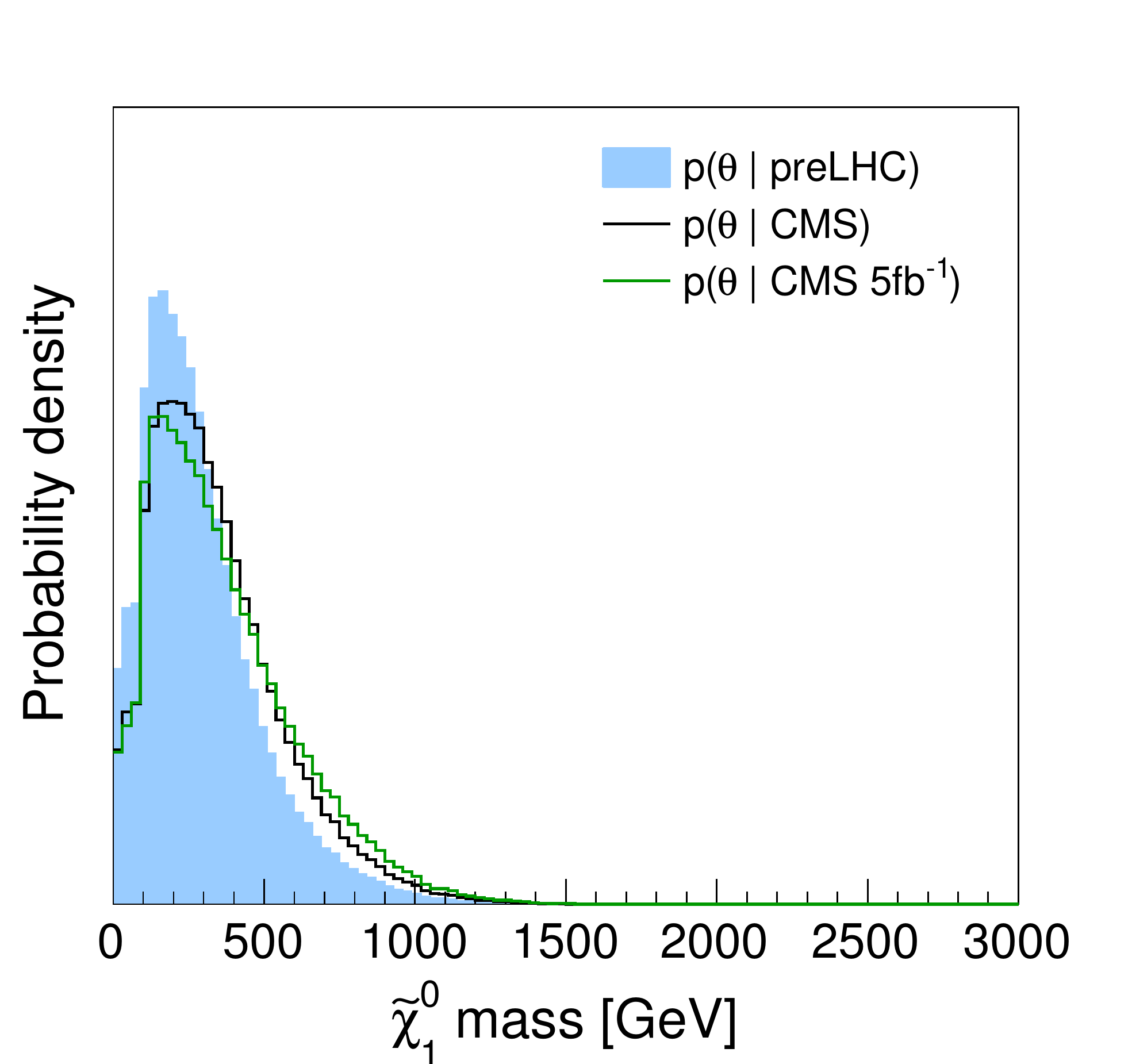} 
   \includegraphics[width=3.6cm]{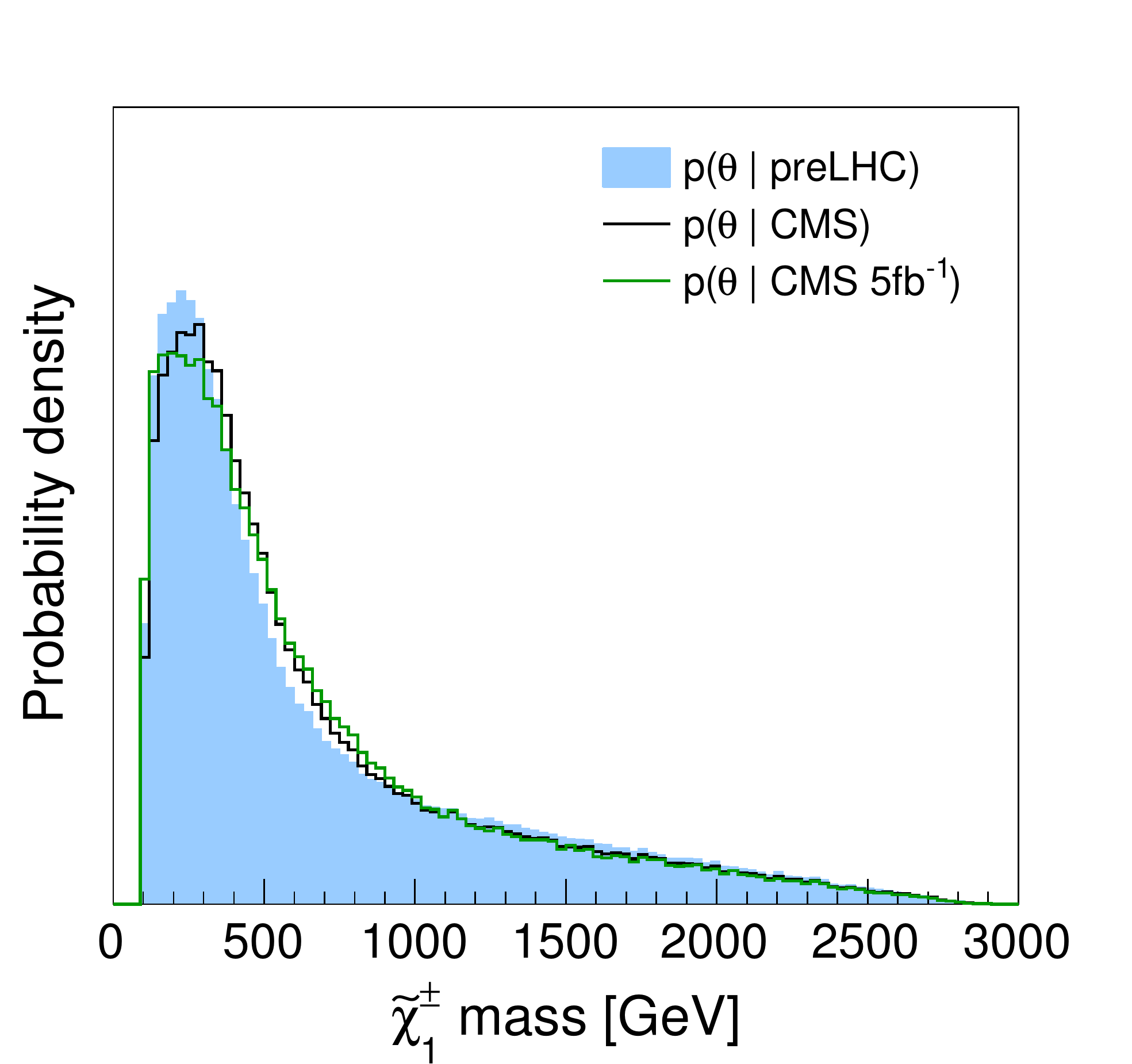} 
   \includegraphics[width=3.6cm]{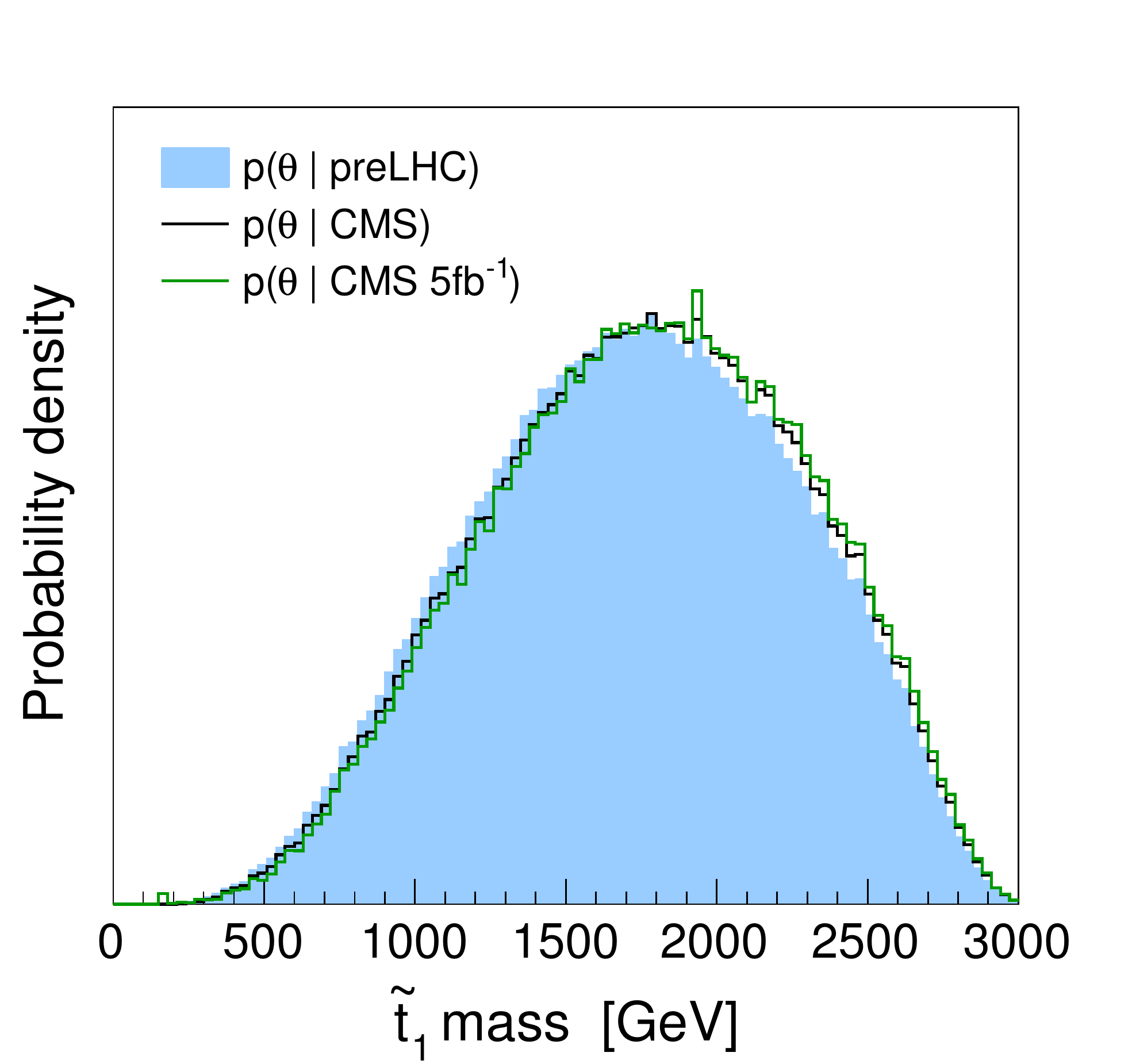} 
   \caption{Marginalized 1D posterior densities of gluino, squark, etc.\ masses comparing the effect of 1~fb$^{-1}$ (black lines) and 5~fb$^{-1}$ (green lines) of data. For the 5~fb$^{-1}$ results, we use simple rescaling.}
   \label{fig:dist-rescale5}
\end{figure}

The effect of going from 1~fb$^{-1}$ to 5~fb$^{-1}$ is illustrated in Fig.~\ref{fig:dist-rescale5}. 
Here we use simple rescaling to obtain results corresponding to 5~fb$^{-1}$, assuming that the analyses do not change too much. As can be seen, the improvement in sensitivity is of the order of 200~GeV for $\tilde g$ and $\tilde q_L$ masses, and a bit more for $\tilde q_R$ masses. The sensitivity to other sparticles hardly improves.

\begin{figure}[t]
   \centering
   \includegraphics[width=5.2cm]{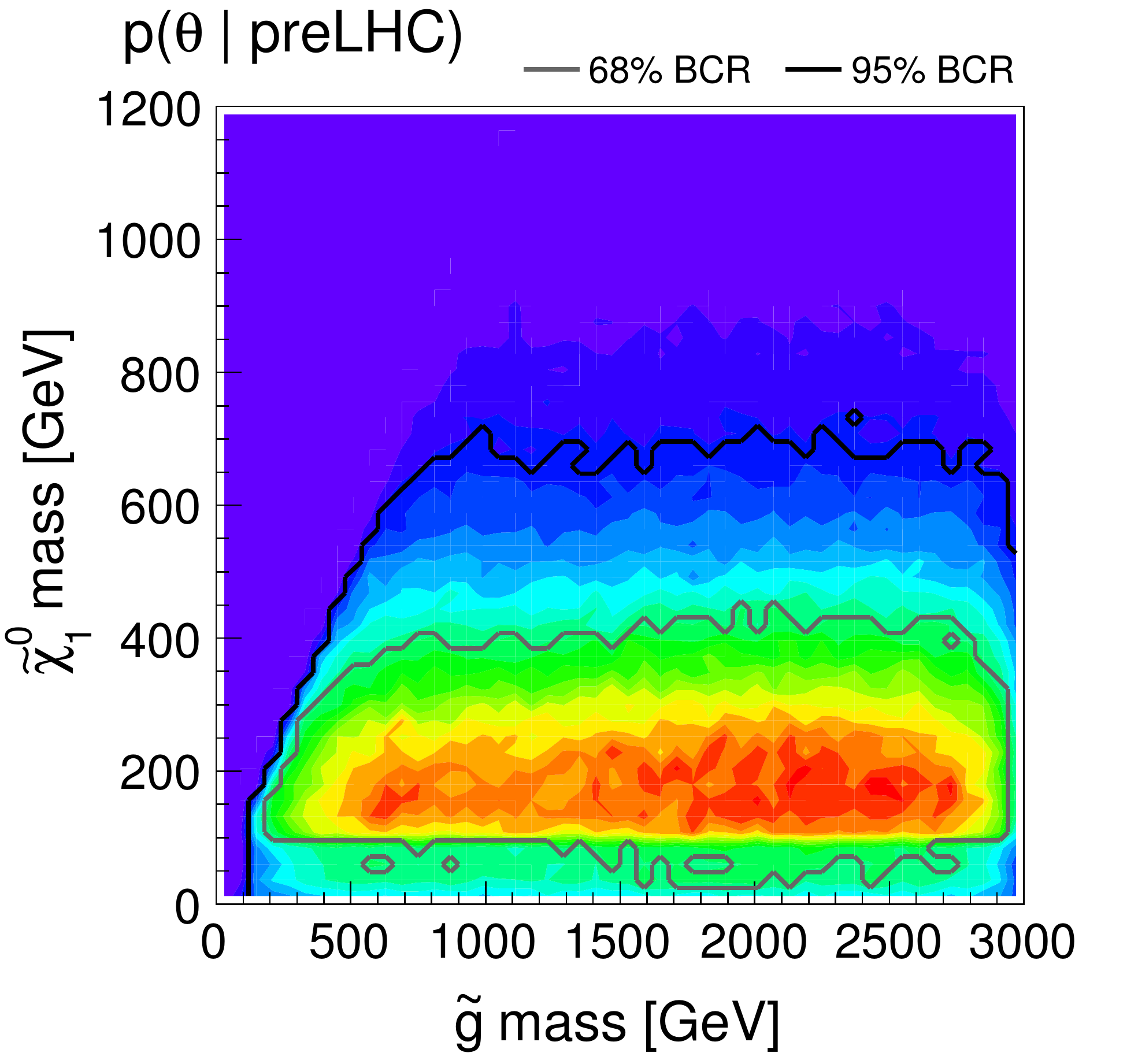}
   \includegraphics[width=5.2cm]{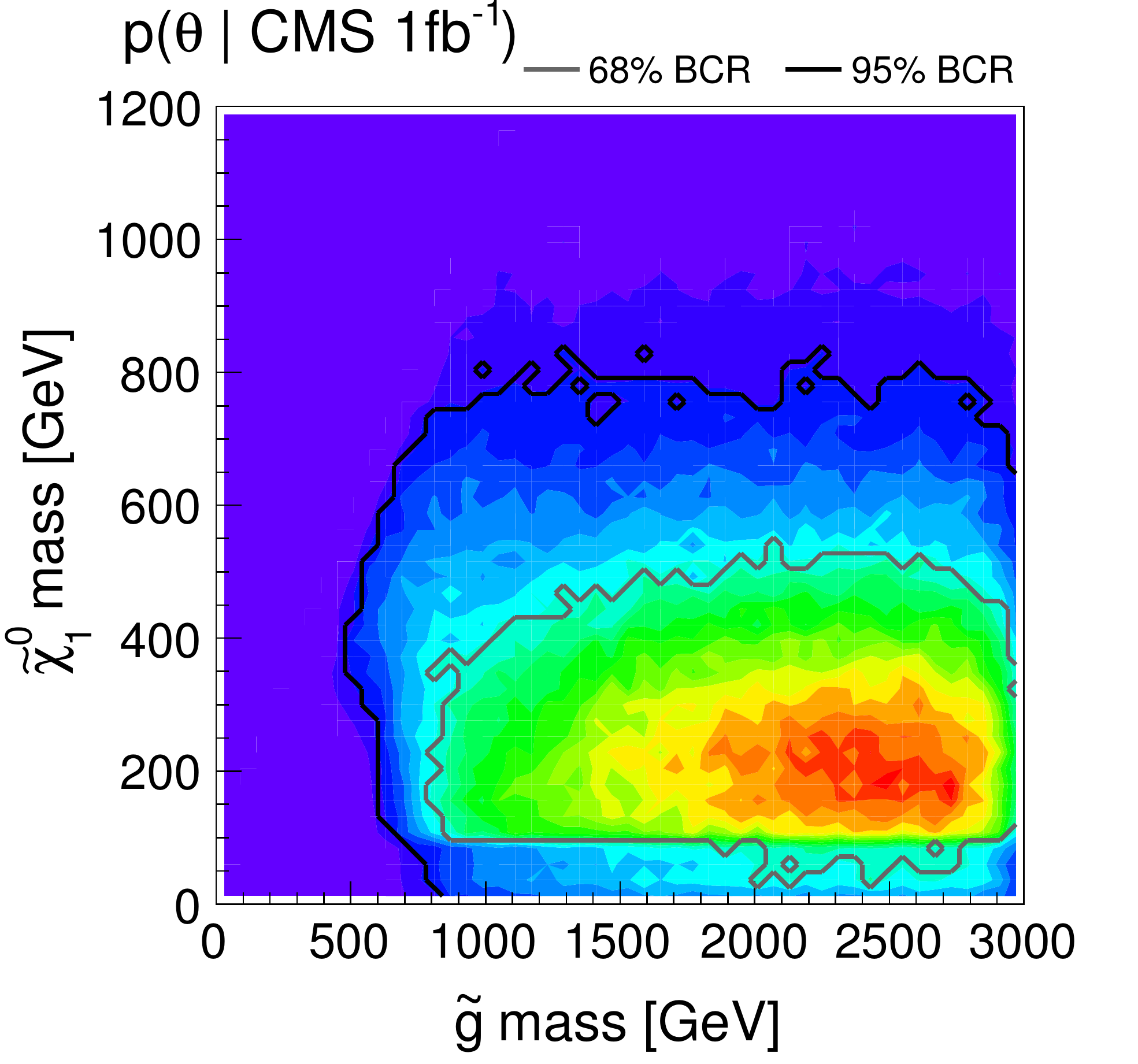}
   \includegraphics[width=5.2cm]{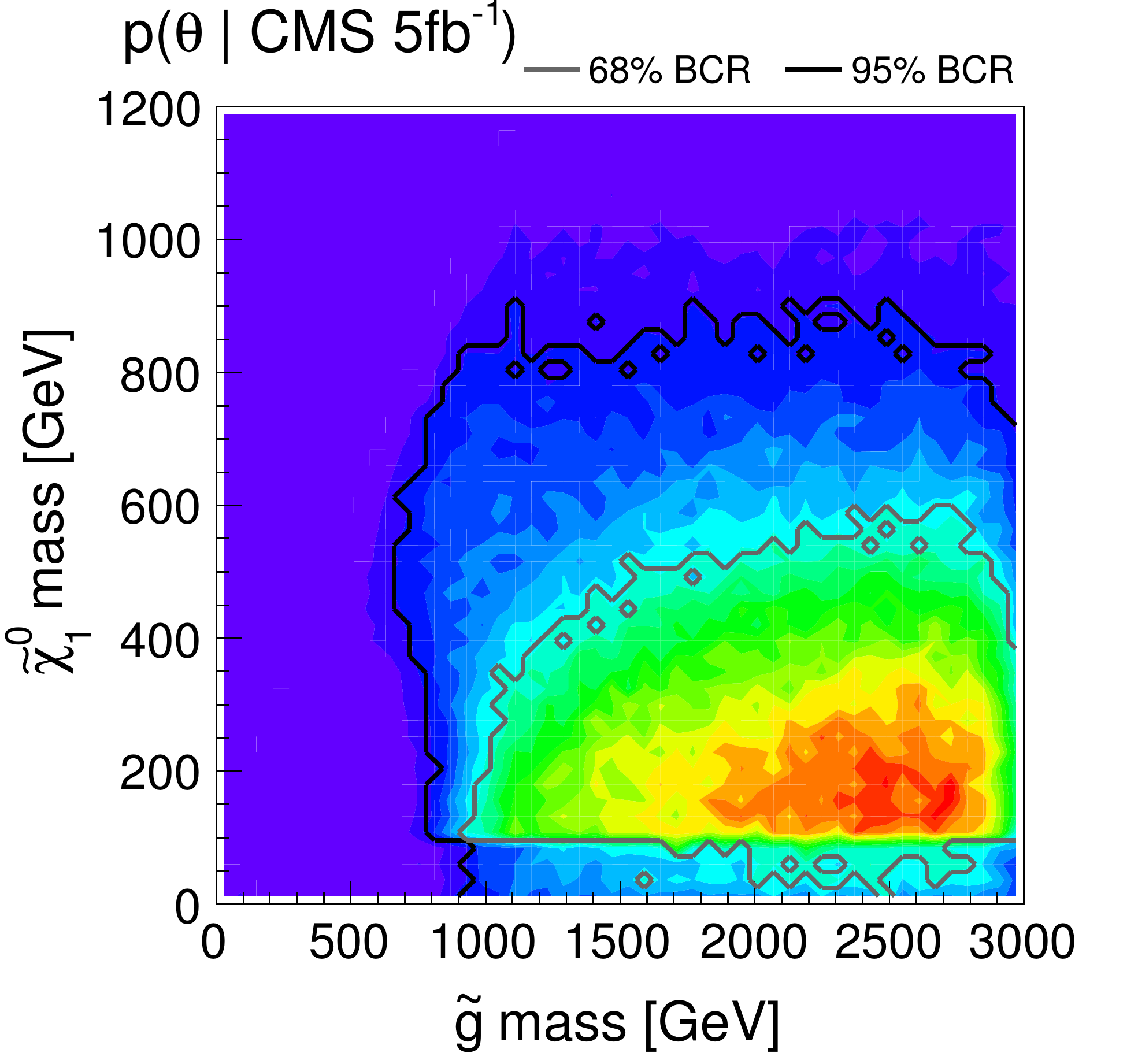} \\[4mm]
   \includegraphics[width=5.2cm]{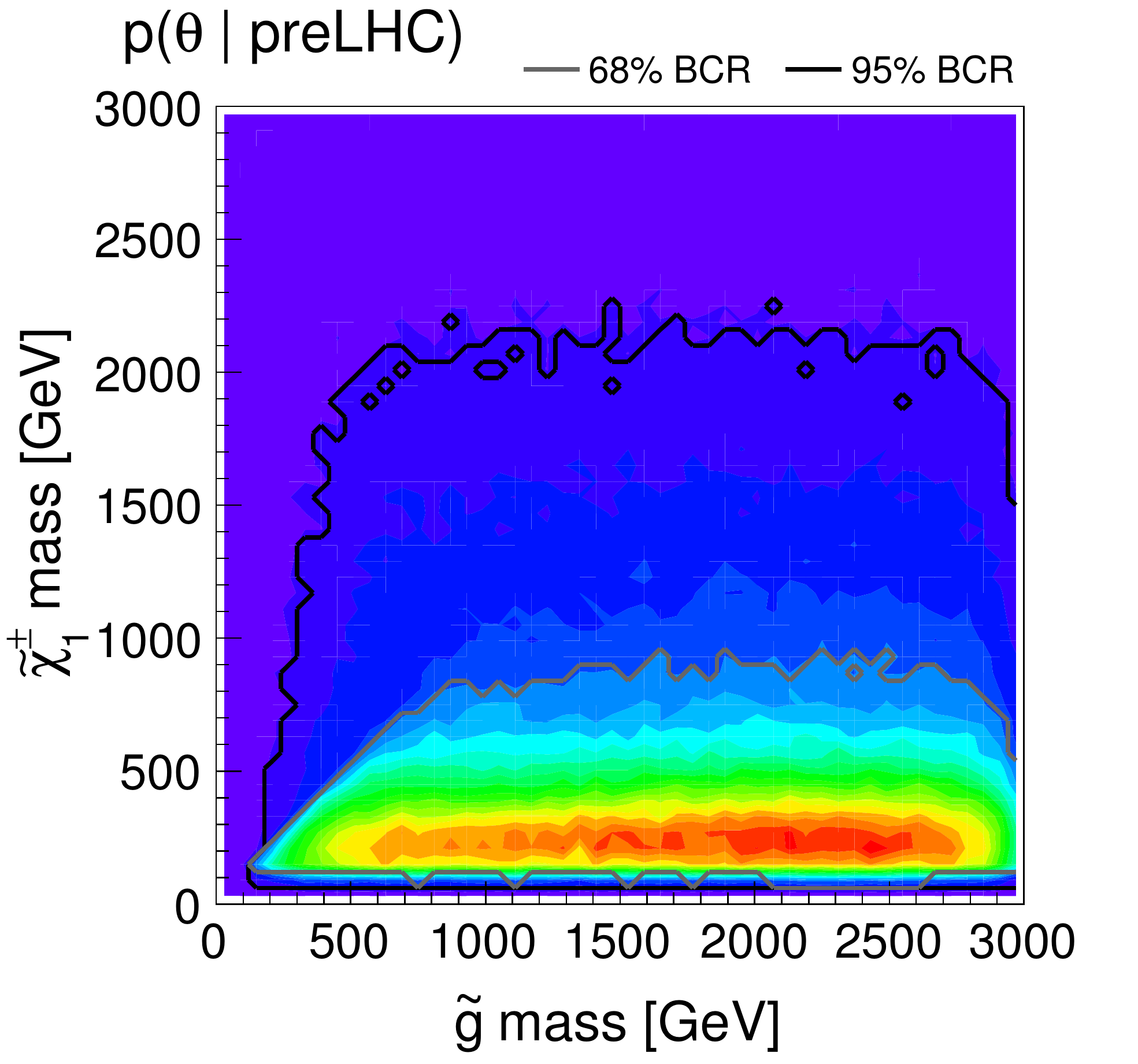}
   \includegraphics[width=5.2cm]{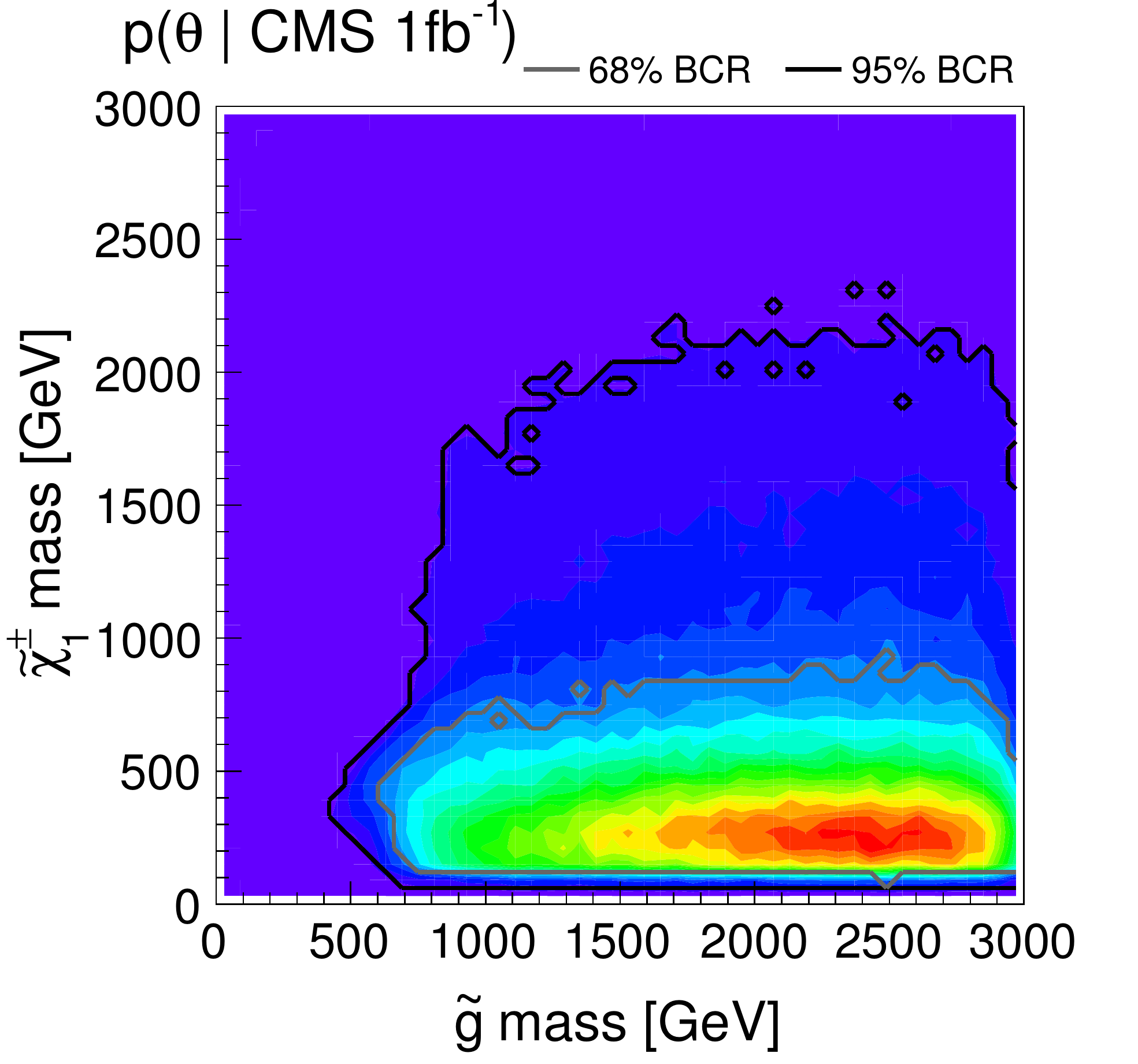}
   \includegraphics[width=5.2cm]{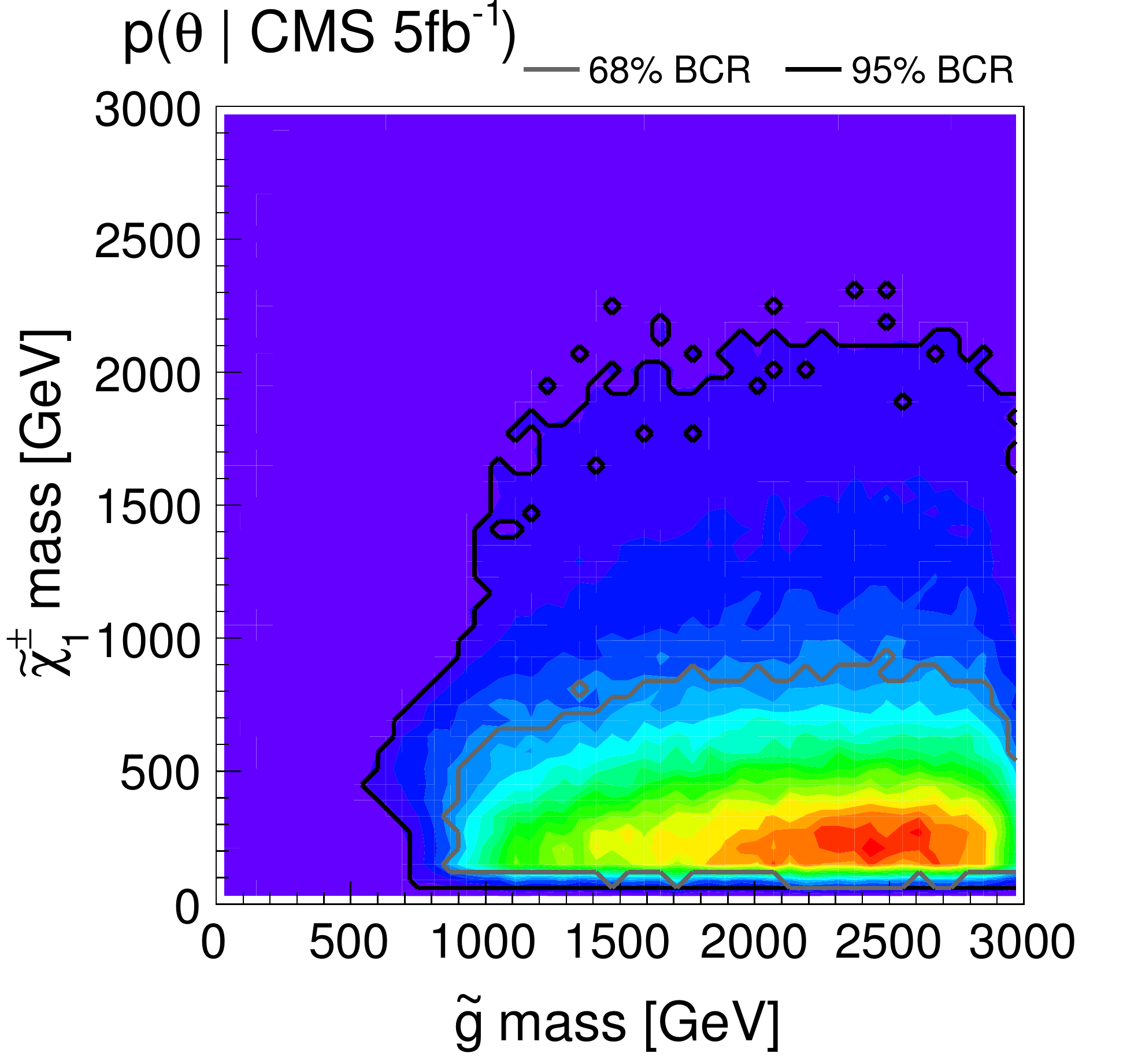} 
   \caption{Marginalized 2D posterior densities of gluino versus neutralino (top row) and of gluino versus chargino (bottom row) mass; on the left preLHC, in the middle for the 1~fb$^{-1}$ analyses, on the right the same rescaled to 5~fb$^{-1}$ of data. The grey and black contours enclose the 68\% and 95\% Bayesian credible regions, respectively.}
   \label{fig:dist-2d-mg-mw1}
\end{figure}

Our approach also allows us to study correlations between sparticle masses  in a straightforward way, 
as illustrated in Fig.~\ref{fig:dist-2d-mg-mw1}. 
As can be seen, the sensitivity to the gluino mass corresponds to that found in the CMSSM only for light charginos and neutralinos. 
As the gluino\,:\,chargino (or gluino\,:\,neutralino) mass ratio decreases, one looses in sensitivity. For $m_{\tilde \chi^0_1},\, m_{\tilde \chi^\pm_1}\gtrsim 400$--500~GeV,  
no limit other than that the gluino must be heavier than the LSP can be derived with 1~fb$^{-1}$. This only marginally changes with 5~fb$^{-1}$ at 7~TeV.

\begin{figure}[t]
   \centering
   \includegraphics[width=12cm]{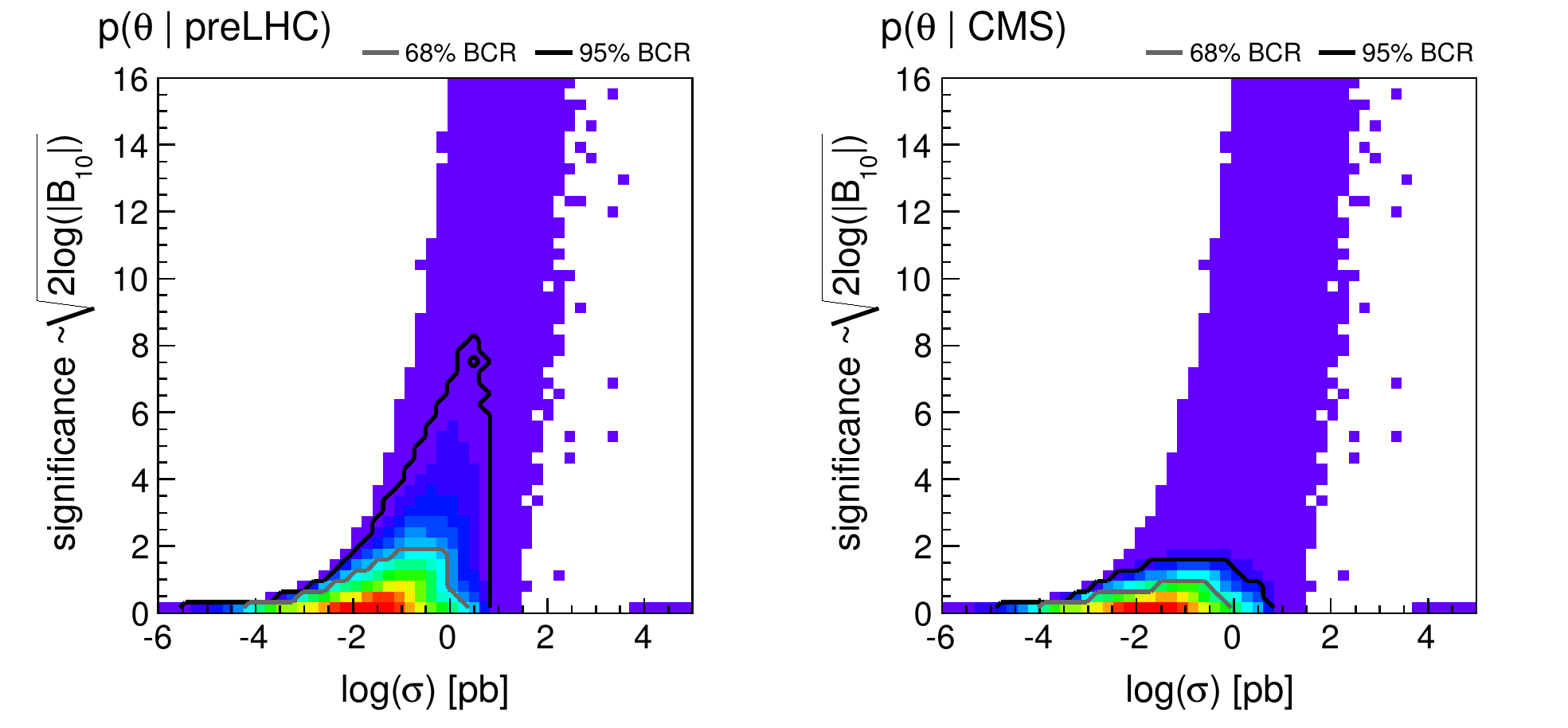}
   \caption{Marginalized 2D posterior densities of signal significance versus total SUSY cross section, on the left before and on the right after taking the CMS searches into account. The grey and black contours enclose the 68\% and 95\% Bayesian credible regions, respectively.}
   \label{fig:dist-2d-signal}
\end{figure}

Let me stress finally that pMSSM points with low signal significance, such that they escape the current searches, do not necessarily have low cross section. This is illustrated in Fig.~\ref{fig:dist-2d-signal}, where we show the signal significance versus total SUSY cross section both preLHC and after the three CMS analyses.  As one can see, points that escape LHC detection (significance $<2$) can still have cross sections as large as $\approx$1~pb. In Fig.~\ref{fig:dist-2d-signal}, about 2/3 of the points with large cross section but low signal significance have dominantly EW-ino production with small $\chi^0_2$\,--\,$\chi^0_1$ mass splitting, so that the decay products are too soft to pass the analysis cuts. Another important class is small $\tilde q_{L,R}^{}$\,--\,$\chi^0_1$ mass splitting, leading to soft jets and low $E_T^{\rm miss}$.

\section{Conclusions}

Current SUSY searches at the LHC are pushing squark and gluino mass limits beyond 1~TeV within constrained models.
However, they still provide rather limited constraints on supersymmetry in general. With the currently available data and searches, we have indeed been able to probe only a small portion of the vast MSSM parameter space, while many regions are still waiting to be explored.  These are in particular ``natural SUSY'' scenarios, scenarios with dominantly EW production, and scenarios with small mass splittings (compressed spectra). There is hence ample of room for SUSY to hide from discovery during the first phase of LHC. 
A complementary  study \cite{Arbey:2011un} based on flat scans comes to analogous conclusions. 

\section*{Acknowledgements}

I wish to thank Sezen Sekmen for a most fruitful and inspiring collaboration during the past five years, and for providing several new plots for this contribution.

\section*{References}

\end{document}